\documentclass[10pt]{iopart}

\usepackage{graphicx} 
\usepackage{cite}     

\begin{document}

\newcommand{\scri}{{\mathcal{I}}}        
\def \d {{\rm d}}

\title[Radiation generated by accelerating and rotating charged black holes]{Radiation generated by accelerating and rotating charged black holes in (anti-)de~Sitter space}

\author{J Podolsk\'y\dag\  and H Kadlecov\'a\ddag}

\address{Institute of Theoretical Physics, Faculty of Mathematics and Physics, Charles University in Prague, 
V~Hole\v{s}ovi\v{c}k\'ach 2, 180~00 Praha 8, Czech Republic }
\eads{\dag\ \mailto{podolskyATmbox.troja.mff.cuni.cz}, \ddag\ \mailto{Hedvika.KadlecovaATcentrum.cz}}

\begin{abstract}  
Asymptotic behaviour of gravitational and electromagnetic fields of exact type~D solutions from the large Pleba\'nski--Demia\'nski family of black hole spacetimes is analyzed. The amplitude and directional structure of radiation is evaluated in cases when the cosmological constant is  non-vanishing, so that the conformal infinities have either de Sitter-like or anti-de Sitter-like character. In particular, explicit relations between the parameters that characterize the sources (that is their mass, electric and magnetic charges, NUT parameter, rotational parameter, and acceleration) and properties of the radiation generated by them are presented. The results further elucidate the physical interpretation of these solutions and may help to understand radiative characteristics of more general spacetimes than those that are asymptotically flat.
\end{abstract}

\submitto{\CQG}

\maketitle

\section{Introduction}

Investigation of various aspects of gravitational waves is one of the main topics of contemporary relativistic physics. In rigorous treatments of gravitational radiation in the framework of Einstein's theory, most studies of this problem have concentrated on radiative spacetimes which are, at least in some directions, asymptotically flat. Until recently, exact ``cosmological'' gravitational waves have attracted less attention. One of the reasons for this is that the definition and characterization of gravitational radiation in such a more general setting is not obvious.

Indeed, standard techniques used for investigation of asymptotically flat spacetimes (such as the Bondi--Sachs approach which allows one to introduce the Bondi mass and momentum, and to characterise the time evolution and energy balance including gravitational radiation in terms of the news  functions --- the analogue of the radiative part of the Poynting vector known from electrodynamics) cannot be applied when the cosmological constant ${\Lambda}$ is non-vanishing. Moreover, even in the relatively simple case of vacuum spacetimes, but with ${\Lambda\not=0}$, it had long been known \cite{Penrose:1964,Penrose:1965,Penrose:1967} that the radiative component of the gravitational field is not unique. It depends substantially on the null direction along which it approaches a given point at conformal infinity~$\scri$ because it has a spacelike character when ${\Lambda>0}$ or timelike character when ${\Lambda<0}$.

It was recently demonstrated in the series of papers \cite{BicakKrtous:2002,BicakKrtous:2005,KrtousPodolsky:2003,PodolskyOrtaggioKrtous:2003,KrtousPodolskyBicak:2003,KrtousPodolsky:2004a,KrtousPodolsky:review,KrtousPodolsky:2005} that such a directional structure of gravitational or electromagnetic radiation can be generally described in a closed form. In fact, it has a universal character that is essentially determined by the algebraic type of the given field, namely by local degeneracy and orientation of its principal null directions on $\scri$. The general behaviour for both the cases ${\Lambda>0}$ and ${\Lambda<0}$ was, respectively, described in \cite{KrtousPodolskyBicak:2003} and~\cite{KrtousPodolsky:2004a}, and subsequently summarized and presented in a unified way in the review \cite{KrtousPodolsky:review}. However, more specific and explicit analyses are highly desirable to exemplify these results. To that end, it is necessary to consider suitable classes of exact radiative solutions of Einstein's equations which admit a nontrivial cosmological constant. A particularly interesting family of such spacetimes is the so called \hbox{C-metric} which can be physically interpreted as describing the field of (possibly charged) non-rotating black holes that uniformly accelerate in the de~Sitter or anti-de~Sitter universe. Detailed studies of the directional structure of radiation and related properties of these spacetimes have already been presented in \cite{KrtousPodolsky:2003,PodolskyOrtaggioKrtous:2003}.

The purpose of the present work is to extend and further generalize these results, namely to consider an effect of a twist. In particular, we will consider the large Pleba\'nski--Demia\'nski family of exact type~D solutions of the Einstein--Maxwell field equations \cite{PlebanskiDemianski:1976} (see also \cite{Debever:1971}). This not only includes  the famous Kerr--Newman rotating black holes, the Taub--NUT spacetime, and the {(anti-)de~Sitter} metric, but also their arbitrary combination. In addition, the  resulting black holes may accelerate due to conical singularities (such as cosmic strings) attached to them. The general form of the metric thus contains seven~free parameters which characterize the mass, charge (electric and magnetic), rotation, the NUT parameter, acceleration of the sources, and the cosmological constant.

This Pleba\'nski--Demia\'nski class of solutions has recently been described in more detail in \cite{GriffithsPodolsky:2005,GriffithsPodolsky:2006a,GriffithsPodolsky:2006b,PodolskyGriffiths:2006}, where a list of further references can also be found. In particular, a new form of the Pleba\'nski--Demia\'nski metric has been presented in which the parameters were given a clear physical interpretation and from which the various well-known special subfamilies can easily be obtained. Two new parameters directly representing the acceleration of the sources and the twist of the repeated principal null congruences have been introduced, and then (following the idea first applied in the particular C-metric cases \cite{HongTeo:2003,HongTeo:2005}) the remaining freedom has been used to simplify the roots of the corresponding metric functions. This has helped significantly in the interpretation of this family of solutions: in particular to distinguish the Kerr-like rotation and the NUT-like contribution to the twist. A simple unified form of the previously known black hole spacetimes has thus also been obtained. We will employ its straightforward coordinate modification in the present analysis.

This paper is organized as follows. In section~2 we will review the Pleba\'nski--Demia\'nski family of solutions, introducing convenient coordinates and frames. In section~3 we will explicitly derive the specific radiative field components of the Pleba\'nski--Demia\'nski spacetimes, using the approach given in~\cite{KrtousPodolsky:2005} which applies to general type~D metrics. We will cover all possible cases which may occur, both for a positive and negative cosmological constant. In the final section~4 we will analyze and visualize the thus obtained directional structure of radiation and, as a main result, its amplitude in various points at conformal infinity.

\section{The general Pleba\'nski--Demia\'nski family of black holes}
\label{genPDfamily}

In \cite{PlebanskiDemianski:1976} Pleba\'nski and Demia\'nski presented a complete class of type~D solutions of Einstein's equations, with a possibly non-zero cosmological constant $\Lambda$, which may be either vacuum or electrovacuum (such that both repeated principal null congruences of the Weyl tensor and that of the non-null electromagnetic field are aligned). As shown in \cite{GriffithsPodolsky:2005}, it is useful to write the general  Pleba\'nski--Demia\'nski metric as
  \begin{equation}
   \hskip-20mm
  \begin{array}{r}
{\displaystyle  \d s^2={\frac{1}{(1-\alpha pr)^2}} \Bigg(
{-Q\over r^2+\omega^2p^2}(\d\tau-\omega p^2\d\sigma)^2 +{\mathcal{P}\over r^2+\omega^2p^2}(\omega\d\tau+r^2\d\sigma)^2 } \hskip1pc \\[12pt]
  {\displaystyle +{r^2+\omega^2p^2\over \mathcal{P}}\,\d p^2 +{r^2+\omega^2p^2\over Q}\,\d r^2 \Bigg),}
  \end{array}
  \label{PleDemMetric}
  \end{equation}
  where ${\mathcal{P}(p)}$ and ${Q(r)}$  are quartic functions
 \begin{equation}
  \hskip-25mm
 \begin{array}{l}
 \hskip10mm {\displaystyle \mathcal{P}=k +2\frac{n}{\omega}p -\epsilon p^2 +2\alpha mp^3 -\Big[\alpha^2(\omega^2 k+e^2+g^2)+\frac{\Lambda}{3}\omega^2\Big]p^4}\,, \\[8pt]
 \hskip10mm Q={\displaystyle(\omega^2k+e^2+g^2) -2mr +\epsilon r^2 -2\alpha\frac{n}{\omega}r^3 -\Big(\alpha^2k+\frac{\Lambda}{3}\Big)\,r^4}\,,
 \end{array}
 \label{PQeqns}
 \end{equation} 
 and $m$, $n$, $e$, $g$, $\Lambda$, $\epsilon$, $k$, $\alpha$ and $\omega$ 
are arbitrary real parameters of which two can be chosen for convenience.\footnote{The original metric given in \cite{PlebanskiDemianski:1976} is recovered by setting ${\,\alpha=1=\omega\,}$ and putting ${\,k=\gamma-g^2-\Lambda/6}$.} Depending on the specific choice of these parameters (and thus the number and types of the roots of the polynomials $\mathcal{P}$ and $Q$) and on the admitted range of the coordinates, the above class of solutions contains a very large number of various spacetimes with different geometrical and physical properties. 

Physically most interesting are those solutions which describe black holes. It was demonstrated in \cite{GriffithsPodolsky:2005,GriffithsPodolsky:2006a,GriffithsPodolsky:2006b,PodolskyGriffiths:2006} that, in such a case, it is particularly convenient to express the parameters $\epsilon$, $n$ and $k$, which occur in the metric functions (\ref{PQeqns}), in  terms of new parameters $a$ and $l$ as
\begin{eqnarray}
\hskip-10mm&& \epsilon=\frac{\omega^2k}{a^2-l^2}+4\alpha\frac{l}{\omega}m-(a^2+3l^2)\,\bigg(\,\frac{\alpha^2}{\omega^2}(\omega^2k+e^2+g^2)+\frac{\Lambda}{3}\,\bigg),\label{p1}\\
\hskip-10mm&& n=\frac{\omega^2k\,l}{a^2-l^2}-\alpha\frac{a^2-l^2}{\omega}m+(a^2-l^2)\,l \bigg(\,\frac{\alpha^2}{\omega^2}(\omega^2k+e^2+g^2)+\frac{\Lambda}{3}\,\bigg),\label{p2}\\
\hskip-10mm&& k=\bigg(1+2\alpha\frac{l}{\omega}m-3\alpha^2\frac{l^2}{\omega^2}(e^2+g^2)-l^2\Lambda\bigg)\bigg(\,\frac{\omega^2}{a^2-l^2}+3\alpha^2l^2\bigg)^{-1}\!\!.\label{p3}
\end{eqnarray}
The remaining scaling freedom can be used to set the twist parameter $\omega$ to {\em any} convenient value, provided $a$ and $l$ do not both vanish. For the particular case ${\,a=0=l\,}$ we apply the relations ${\,\epsilon=1-\alpha^2(e^2+g^2)\,}$, ${\,n=0\,}$, ${\,k=1}$.

We are thus left with seven physical parameters, namely ${\Lambda, m, e, g, a, l}$ and~$\alpha$. These have direct interpretation as representing the cosmological constant, mass parameter, electric charge, magnetic charge, Kerr-like rotation, NUT parameter and the acceleration of the black holes, respectively. Indeed, performing the simple transformation
  \begin{equation}
  p=\frac{l}{\omega}+\frac{a}{\omega}\cos\theta\,, \qquad  
  \tau=t-\frac{(l+a)^2}{a}\,\phi\,, 
  \qquad \sigma=-\frac{\omega}{a}\,\phi\,,
  \label{trans1A}
  \end{equation}
the Pleba\'nski--Demia\'nski metric (\ref{PleDemMetric}), (\ref{PQeqns}) becomes 
  \begin{equation}
  \hskip-20mm
  \begin{array}{l}
{\displaystyle \d s^2=\frac{1}{\>\hat\Omega^2}\,\bigg\{
-\frac{Q}{\>\rho^2}\Big[\d t- \Big(a\,\sin^2\theta +4l\,\sin^2\!\frac{\,\theta}{2}\, \Big)\d\phi \Big]^2  +  \frac{\>\rho^2}{Q}\,\d r^2 
} \\[8pt]
  \hskip6pc {\displaystyle 
  +\frac{\>\rho^2}{P}\,\d\theta^2  + \frac{P}{\>\rho^2} \sin^2\theta \Big[\, a\,\d t+\Big(r^2+(a+l)^2\Big)\d\phi\, \Big]^2  
\bigg\}, }
\end{array}
  \label{newMetric}
  \end{equation}
  where 
  \begin{equation}
  \begin{array}{l}
  {\displaystyle \hat\Omega=1-{\alpha\over\omega}(l+a\cos\theta)\,r\,, } \\[6pt]
  \rho^2 =r^2+(l+a\cos\theta)^2\,, \\[6pt]
  P= 1-a_3\cos\theta-a_4\cos^2\theta\,, \\
  Q= {\displaystyle (\omega^2k+e^2+g^2) -2mr +\epsilon \,r^2 
-2\alpha\frac{n}{\omega}\, r^3 -\Big(\alpha^2k+\frac{\Lambda}{3}\,\Big)\,r^4,} 
  \end{array} 
  \label{newMetricFns}
  \end{equation} 
  and 
 \begin{equation}
 \begin{array}{l}
  {\displaystyle a_3= 2\alpha{a\over\omega}m -4\alpha^2{al\over\omega^2} (\omega^2k+e^2+g^2) -4{\Lambda\over3}al }\,, \\[6pt]
  {\displaystyle a_4= -\alpha^2{a^2\over\omega^2}(\omega^2k+e^2+g^2) -{\Lambda\over3}a^2\,, }
  \end{array}
 \label{a34}
 \end{equation} 
with $\epsilon$, $n$ and $k$ in $Q$ given by (\ref{p1})--(\ref{p3}). It is also assumed that $|a_3|$ and $|a_4|$ are sufficiently small that $P$ has no roots within the considered range ${\theta\in[0,\pi]}$. 

This general metric immediately reduces to the Kerr--Newman--NUT--de~Sitter solution when the acceleration vanishes, ${\alpha=0}$, and further to the familiar forms of the Kerr--Newman--de~Sitter black hole spacetimes when the NUT parameter~$l$ is set to zero or to the charged NUT--de~Sitter spacetime when the rotational Kerr-like parameter~$a$ is set to zero. Then, putting the charges $e$, $g$ and the cosmological constant $\Lambda$ to zero, the well-known forms of Kerr (and Schwarzschild) and Taub-NUT metrics are recovered. On the other hand, when ${\alpha\not=0}$ but we set ${l=0}$ and then ${a=\omega=0}$, the non-twisting C-metric solution is immediately obtained which describes charged black holes uniformly accelerating in Minkowski, de~Sitter or anti-de~Sitter universe (according to the sign of $\Lambda$). Therefore, the line element (\ref{newMetric}) is a very convenient metric representation of the complete class of accelerating, rotating and charged black holes of the Pleba\'nski--Demia\'nski class. 

Let us also recall that the spacetime has a Kerr-like ring singularity at ${\rho=0}$ when ${|a|\geq |l|}$. When ${|a|<|l|}$, the metric is singularity free. Although these two cases have identical metric forms (\ref{newMetric}), their singularity and global structures differ substantially.

\subsection{Alternative form of the metric}

For an investigation of the asymptotic behaviour of gravitational and electromagnetic radiation it is necessary to cover completely the regions near the conformal infinity~$\scri$. However, for the Pleba\'nski--Demia\'nski family of space-times this is difficult to achieve using the coordinates of the above metric (\ref{newMetric}). The main reason is that the coordinate $r\in(0,+\infty)$ does not extend up to $\scri\,$, which is located at ${\,\hat\Omega=0\,}$, in \emph{all} directions, that is for all values of $\theta$. It will thus be convenient to introduce another metric form of the Pleba\'nski--Demia\'nski family of black holes. This is obtained from (\ref{PleDemMetric}) by introducing the reciprocal radial coordinate ${\,q=-r^{-1}}$. The resulting metric\footnote{In fact, for ${\alpha=1=\omega}$ this reduces to the original form of the Pleba\'nski--Demia\'nski metric given in \cite{PlebanskiDemianski:1976}, see (2.1) and (3.3) therein.} reads
\begin{equation}
\hskip-20mm
\d s^2={\frac{1}{\Omega^2}} \bigg(\!-\frac{\mathcal{Q}}{\>\varrho^2}\,(\d\tau-\omega p^2\d\sigma)^2 
+\frac{\mathcal{P}}{\>\varrho^2}\,(\omega q^2\d\tau+\d\sigma)^2 +\frac{\>\varrho^2}{\mathcal{P}}\,\d p^2 + \frac{\>\varrho^2}{\mathcal{Q}}\,\d q^2 \bigg),
  \label{PleDemMetricNEW}
\end{equation}
where 
\begin{equation}
  \hskip-28mm
 \begin{array}{l}
 \hskip10mm \Omega\>\>=\>-(q+\alpha p)\,, \\[7pt]
 \hskip10mm \varrho^2\>=\>1+\omega^2p^2q^2\,, \\[5pt]
 \hskip10mm \mathcal{P}(p)={\displaystyle k +2\frac{n}{\omega}p -\epsilon p^2 +2\alpha mp^3 -\Big[\alpha^2(\omega^2 k+e^2+g^2)+\frac{\Lambda}{3}\omega^2\Big]p^4}\,, \\[6pt]
 \hskip10mm \mathcal{Q}(q)={\displaystyle-\Big(\alpha^2k+\frac{\Lambda}{3}\Big) +2\alpha\frac{n}{\omega}q+\epsilon q^2 + 2m q^3 + (\omega^2k+e^2+g^2)\,q^4}\,.
 \end{array}
 \label{PQeqnsNEW}
\end{equation} 
The parameters  $\epsilon$, $n$ and $k$ are still given by (\ref{p1})--(\ref{p3}). The roots of $\mathcal{Q}$ given by ${\mathcal{Q}(q)=0}$ represent horizons at specific values of ${\,q<0\,}$. To retain a Lorentzian signature of the metric (\ref{PleDemMetricNEW}) it is necessary that ${\mathcal{P}>0}$. The coordinate $p$ must thus be restricted to a particular range between appropriate roots ${\mathcal{P}=0}$ which correspond to poles of the black holes. This is automatically achieved by assuming the parameterization ${\,p=(l/\omega)+(a/\omega)\cos\theta\,}$ with ${\,\theta\in[0,\pi]}$. In fact, to express all Pleba\'nski--Demia\'nski black hole spacetimes --- including those with a non-trivial NUT parameter $l$ --- in a closed and explicit form it is useful to finally apply the specific linear transformation (\ref{trans1A}) on the metric (\ref{PleDemMetricNEW}). This would yield the metric analogous to (\ref{newMetric}), only with the ``radial'' coordinate $r$ replaced by $q$ via the above relation ${\,q=-r^{-1}}$. Notice also that, consequently,  ${\,\Omega>0\,}$ in the whole spacetime.

In order to characterize the directional structure and asymptotic behaviour of radiation, it is first necessary to express the Weyl and the electromagnetic tensor in the natural frame of null vectors ${{\bf k},{\bf l},{\bf m},{\bf \bar{m}}}$, such that ${\,{g}_{\alpha\beta}\,{k}^{\alpha}{l}^{\beta}=-1\,}$, ${\,{g}_{\alpha\beta}\,{m}^{\alpha}{\bar{m}}^{\beta}=+1\,}$. We need to distinguish two possible situations, in which the region near the conformal infinity $\scri$ is either stationary (${\mathcal{Q}>0}$) or non-stationary (${\mathcal{Q}<0}$).

In the regions when ${\mathcal{Q}>0}$, the null frame is
\begin{eqnarray}
{\bf k}&=&\frac{{\Omega}}{\sqrt{2}\,\varrho}
\left(\frac{1}{\sqrt{\mathcal{Q}}}\,(\,\partial_{\tau}-\omega q^2\partial_{\sigma } )+\sqrt{\mathcal{Q}}\,\partial_{q}\right),\nonumber\\
{\bf l}&=&\frac{{\Omega}}{\sqrt{2}\,{\varrho}}
\left(\frac{1}{\sqrt{\mathcal{Q}}}\,(\,\partial_{\tau}-\omega q^2\partial_{\sigma } )-\sqrt{\mathcal{Q}}\,\partial_{q}\right),\nonumber\\
{\bf m}&=&\frac{{-\Omega}}{\sqrt{2}\,{\varrho}}
\left(\sqrt{\mathcal{P}}\,\partial_{p}+\frac{\hbox{i}}{\sqrt{\mathcal{P}}}\,(\,\omega p^2\partial_{\tau }+\partial_{\sigma})\right),\label{altervectorPLUS}\\
{\bf \bar{m}}&=&\frac{{-\Omega}}{\sqrt{2}\,{\varrho}}
\left(\sqrt{\mathcal{P}}\,\partial_{p}-\frac{\hbox{i}}{\sqrt{\mathcal{P}}}\,(\,\omega p^2\partial_{\tau}+\partial_{\sigma})\right).\nonumber
\end{eqnarray}

In the complementary case ${\mathcal{Q}<0}$, the null frame is
\begin{eqnarray}
{\bf k}&=&\frac{{\Omega}}{\sqrt{2}\,\varrho}
\left(\sqrt{-\mathcal{Q}}\,\partial_{q}-\frac{1}{\sqrt{-\mathcal{Q}}}\,(\,\partial_{\tau}-\omega q^2\partial_{\sigma } )\right),\nonumber\\
{\bf l}&=&\frac{{\Omega}}{\sqrt{2}\,{\varrho}}
\left(\sqrt{-\mathcal{Q}}\,\partial_{q}+\frac{1}{\sqrt{-\mathcal{Q}}}\,(\,\partial_{\tau}-\omega q^2\partial_{\sigma } )\right),\nonumber\\
{\bf m}&=&\frac{{\Omega}}{\sqrt{2}\,{\varrho}}
\left(\sqrt{\mathcal{P}}\,\partial_{p}-\frac{\hbox{i}}{\sqrt{\mathcal{P}}}\,(\,\omega p^2\partial_{\tau }+\partial_{\sigma})\right),\label{altervectorMINUS}\\
{\bf \bar{m}}&=&\frac{{\Omega}}{\sqrt{2}\,{\varrho}}
\left(\sqrt{\mathcal{P}}\,\partial_{p}+\frac{\hbox{i}}{\sqrt{\mathcal{P}}}\,(\,\omega p^2\partial_{\tau}+\partial_{\sigma})\right).\nonumber
\end{eqnarray}
The only non-zero component of the Weyl tensor in both these frames is 
\begin{equation}
{\Psi }_2=\left[(m+\hbox{i}\,n)+(e^2+g^2)\left(\frac{q-\alpha p}{1+\hbox{i}\,\omega pq}\right)\right]\left(\frac{q+\alpha p }{1-\hbox{i}\,\omega pq}\right)^3,
\label{weyl1alter}
\end{equation}
which explicitly demonstrates that the spacetimes studied are of algebraic type D and that the (future oriented) tetrad vectors ${\,{\bf k}\,}$ and  ${\,{\bf l}\,}$ are aligned with the double degenerate principal null directions of the Weyl tensor.  The only non-zero component of the Ricci tensor associated with the aligned electromagnetic field is
\begin{equation}
{\rm \Phi}_{11}=\frac{1}{2}(e^2+g^2)\frac{\Omega^4}{\varrho^4}\,,
\label{riccialter2}
\end{equation}
which shows that for ${\,e=0=g\,}$ the spacetimes are vacuum. In fact, the corresponding vector potential is
\begin{equation}
\hbox{A}=\frac{e\,q\,(\,\d \tau - \omega p^2\d\sigma )-g\,p\, (\,\d\sigma+\omega q^2\d \tau)}{1+\omega^2p^2q^2}\,,
\label{vector potential}
\end{equation}
so that the electromagnetic field ${\hbox{F}=\d\hbox{A}}$ is given by
\begin{eqnarray}
&&\hskip-25.5mm\hbox{F}=
\varrho^{-4}\Big\{[e(1-\omega^2p^2q^2) - 2g \omega pq   ]\,\d q\wedge\d \tau
-[g(1-\omega^2p^2q^2) + 2e \omega pq ]\,\d p \wedge\d\sigma
\label{elmagfield}\\
&&\hskip-20.5mm
-\omega p^2[e(1-\omega^2p^2q^2) - 2g \omega pq   ]\,\d q\wedge\d \sigma
-\omega q^2[g(1-\omega^2p^2q^2) + 2e \omega pq   ]\,\d p \wedge\d\tau\Big\}
 \,.\nonumber
\end{eqnarray}
With respect to the null frame (\ref{altervectorPLUS}) we thus obtain
\begin{equation}
{\rm \Phi}_{1}=
\frac{\Omega^2}{2\varrho^4}\Big\{ e(1-\omega^2p^2q^2) - 2g \omega pq
-\hbox{i}\,[g(1-\omega^2p^2q^2) + 2e \omega pq   ] \Big\},
\label{Phi1explicit}
\end{equation}
while with respect to the frame (\ref{altervectorMINUS}) we obtain the complex conjugate of (\ref{Phi1explicit}). In both these cases the relation ${{\rm \Phi}_{11}=2\,{\rm \Phi}_1{\rm \bar\Phi}_1}$ yields (\ref{riccialter2}).

\section{Asymptotic structure of radiation}
\label{radiation}

We will now investigate the behaviour of radiation in the above family of solutions.

\subsection{The conformal infinity}
\label{sec:coninfty}

The first step is to identify the conformal infinity $\scri$ of the spacetimes. It turns out that the above metric (\ref{PleDemMetricNEW}) is particularly convenient because it is naturally related to a suitable conformal (unphysical) metric $\tilde{g}_{\alpha\beta}$ through the conformal factor ${\Omega>0}$,
\begin{equation}
\tilde{g}_{\alpha\beta}=\Omega^2{g}_{\alpha\beta}\,,\qquad \Omega=-\left(q+\alpha p\right).
\label{metrics}
\end{equation}
The conformal infinity $\scri\,$ is located at ${\Omega=0}$ which obviously occurs at
\begin{equation}
 \scri\,:\qquad q=-\alpha p\,,\label{skraj}
\end{equation}
where the conformal metric $\tilde{g}_{\alpha\beta}$ evaluated on scri, that is $\tilde{{g}}_{\alpha\beta}|_{\mathcal{I}}$, is regular. Since $p$ is finite, $\scri\,$ occurs at finite $q$, so that the conformal infinity of all Pleba\'nski--Demia\'nski black hole space-times is completely covered. In particular, in the equatorial plane given by ${p=0\,}$ it is simply located at ${q=0}$. This parameterisation is also well behaved in the limit with vanishing acceleration (${\alpha=0}$), in which case the whole scri is given by ${q=0}$, that is ${r=\infty}$. This is a standard result for Kerr--Newman--(anti-)de Sitter black holes.

It is then necessary to find the vector ${\tilde{\bf n}}$ normal to conformal infinity $\scri$. This is defined as ${\,\tilde{n}^{\alpha}=\tilde{N}\,\tilde{g}^{\alpha\beta}\,\d_\beta\Omega\,}$ where ${\tilde{N}}$ is a suitable function, cf. \cite{KrtousPodolsky:review}. Since the gradient of ${\,\Omega\,}$ is ${\,\d\Omega=-(\d{q}+\alpha {\d}p)}$, it follows that ${\tilde{\bf n}=-\tilde{N}\varrho^{-2}(\mathcal{Q}\,\partial_{q}+\alpha\mathcal{P}\,\partial_{p})}$. To set the appropriate normalization ${\>\tilde{g}_{\alpha\beta}\,\tilde{n}^{\alpha}\tilde{n}^{\beta}=-\mbox{sign}\,\Lambda\>}$ on conformal infinity, it is necessary to choose ${\tilde{N}^2=\frac{3}{|\Lambda|}}$. Indeed, the vector 
\begin{equation}
\tilde{\bf n}=-\sqrt{\frac{3}{|\Lambda|}}\ \varrho^{-2}(\mathcal{Q}\,\partial_{q}+\alpha\mathcal{P}\,\partial_{p})
\label{normal to scri}
\end{equation}
normal to $\,\scri\,$ has the norm ${\,\tilde{g}_{\alpha\beta}\,\tilde{n}^{\alpha}\tilde{n}^{\beta}=\frac{3}{|\Lambda|}\ \varrho^{-2}(\mathcal{Q}+\alpha^2\mathcal{P})\,}$, in which the function
\begin{eqnarray}
 \hskip-24mm
\mathcal{Q}+\alpha^2\mathcal{P}=-\frac{\Lambda}{3}(1+\alpha^2\omega^2p^4)&-&\Omega\,\Big[(\omega^2k+e^2+g^2)(q-\alpha p)(q^2+\alpha^2p^2)\\
    &&\hskip-2mm +2m(q^2-\alpha qp+\alpha^2p^2)+\epsilon(q-\alpha p)+2\alpha (n/\omega)\Big]\,,\nonumber\label{hura}
\end{eqnarray}
evaluated on scri (where ${\Omega=0}$) obviously reads
\begin{equation}
\left(\mathcal{Q}+\alpha^2\mathcal{P}\right)_{\scri}=-\frac{\Lambda}{3}\left(1+\alpha^2\omega^2 p^4\right)=-\frac{\Lambda}{3}\varrho^2_{\,\scri}\,.\label{nice}
\end{equation}
Therefore, on $\scri$ it follows that ${\,\left(\tilde{g}_{\alpha\beta}\,\tilde{n}^{\alpha}\tilde{n}^{\beta}\right)_{\scri}=-1}$ or $+1$ when ${\Lambda>0}$ or ${\Lambda<0}$, respectively. Thus, in the asymptotically de~Sitter spacetimes ${\tilde{\bf n}}$ is the unit timelike vector, while in the asymptotically anti-de~Sitter spacetimes it is spacelike, as required.

The relation (\ref{nice}) also implies ${\,\mathcal{Q}_{\scri}=-\frac{\Lambda}{3}-\alpha^2\left(\mathcal{P}(p)+\frac{\Lambda}{3}\omega^2 p^4\right)_{\scri}\,}$. Moreover,
\begin{equation}
\mathcal{P}(p)>0\quad\hbox{everywhere\,}.\label{niceplus}
\end{equation}
For ${\Lambda>0}$ the function $\mathcal{Q}$ is thus negative at all points on $\scri$, which means that the complete region close to the conformal infinity is non-stationary. In the case ${\Lambda<0}$ the situation is more complicated: the regions where ${\mathcal{Q}>0}$ are stationary while the alternative regions where ${\mathcal{Q}<0}$ are non-stationary, and these are separated by the Killing horizons located at ${\mathcal{Q}=0}$. This structure occurs also at the conformal infinity (which has the anti-de~Sitter-like character). For small values of the acceleration parameter $\alpha$ all regions of $\scri$ are stationary while for large acceleration the non-stationary region are also present:     
\begin{eqnarray}
\mathcal{Q}_{\scri} > 0 && \qquad \hbox{iff} \qquad \alpha^2\Big(\mathcal{P}(p)+\frac{\Lambda}{3}\omega^2 p^4\Big) <-\frac{\Lambda}{3} \ ,\label{znQ>0}\\
\mathcal{Q}_{\scri} < 0 && \qquad \hbox{iff} \qquad \alpha^2\Big(\mathcal{P}(p)+\frac{\Lambda}{3}\omega^2 p^4\Big) >-\frac{\Lambda}{3} \ .
\label{znQ<0}
\end{eqnarray}
In the equatorial section where ${p=0}$, these conditions simply reduce to ${\alpha^2k<-\frac{\Lambda}{3}}$ and ${\alpha^2k>-\frac{\Lambda}{3}}$, respectively.

\subsection{The structure of radiation when ${\Lambda>0}$}
\label{sec:deSitterradiation}

We will first derive the explicit form of the field components which describe the radiation in the Pleba\'nski--Demia\'nski family of spacetimes in the case when the cosmological constant is positive. In such a case the conformal infinity $\,\scri\,$ has a de~Sitter-like character with the unit normal vector (\ref{normal to scri}) being timelike. 

The function $\mathcal{Q}$ is \emph{everywhere negative} near and on such scri, so that we will employ the null frame ${{\bf k},{\bf l},{\bf m},{\bf \bar{m}}}$ given by (\ref{altervectorMINUS}), which is adapted to both principal null directions. Since ${{g}_{\alpha\beta}\,{k}^{\alpha}{n}^{\beta}=-\sqrt{\frac{3}{\Lambda}}\,\sqrt{-\frac{1}{2}\mathcal{Q}}\,\varrho^{-1}={g}_{\alpha\beta}\,{l}^{\alpha}{n}^{\beta}}$,  where ${{\bf n}=\Omega\,\tilde{\bf n}}$, the frame is future-oriented and it also satisfies the normalization condition ${\epsilon_1=\epsilon_2}$, see equations (18) and (19) in \cite{KrtousPodolsky:2005}. The associated orthonormal tetrad 
${{\bf t}_{\rm s},{\bf q}_{\rm s},{\bf r}_{\rm s},{\bf s}_{\rm s}}$:
\begin{equation}
{\bf k}={\textstyle\frac{1}{\sqrt{2}}}\left({\bf t}_{\rm s}+{\bf q}_{\rm s}\right),\quad
{\bf l}={\textstyle\frac{1}{\sqrt{2}}}\left({\bf t}_{\rm s}-{\bf q}_{\rm s}\right),\quad
{\bf m}={\textstyle\frac{1}{\sqrt{2}}}\left({\bf r}_{\rm s}-\hbox{i}\,{\bf s}_{\rm s}\right),\label{orntonullvector}
\end{equation}
adapted to these algebraically special null directions, is thus
\begin{equation}
 \begin{array}{l}
 {\bf t}_{\rm s}={\displaystyle  \frac{\Omega}{\varrho}\sqrt{{-\mathcal{Q}}}\,\partial_{q}}\,, \\[8pt]
 {\bf q}_{\rm s}={\displaystyle  \frac{\Omega}{\varrho}\frac{-1}{\sqrt{-\mathcal{Q}}}\,(\,\partial_{\tau}-\omega q^2\partial_{\sigma} )}\,, \\[8pt]
 {\bf r}_{\rm s}={\displaystyle  \frac{\Omega}{\varrho}\sqrt{\mathcal{P}}\,\partial_{p}}\,, \\[8pt]
 {\bf s}_{\rm s}={\displaystyle  \frac{\Omega}{\varrho}\frac{1}{\sqrt{\mathcal{P}}}\,(\,\omega p^2\partial_{\tau}+\partial_{\sigma})}\,.
 \end{array}
 \label{sp+}
\end{equation} 
Now, it is natural to choose the following reference orthonormal tetrad adapted to $\scri$:
\begin{equation}
 \begin{array}{l}
 {\bf t}_{\rm o}={\displaystyle  \frac{\Omega}{\varrho}\frac{-1}{\sqrt{{-\left(\mathcal{Q}+\alpha^2\mathcal{P}\right)}}}
\,(\mathcal{Q}\,\partial_{q}+\alpha\mathcal{P}\,\partial_{p})}\,, \\[8pt]
 {\bf q}_{\rm o}={\displaystyle  \frac{\Omega}{\varrho}\frac{-1}{\sqrt{-\mathcal{Q}}}\,(\,\partial_{\tau}-\omega q^2\partial_{\sigma} )}\,, \\[8pt]
 {\bf r}_{\rm o}={\displaystyle  \frac{\Omega}{\varrho}\sqrt{ \frac{\mathcal{P}\mathcal{Q}}{\mathcal{Q}+\alpha^2\mathcal{P}}  }\,(\,\partial_{p}-\alpha \partial_{q})}\,, \\[8pt]
 {\bf s}_{\rm o}={\displaystyle  \frac{\Omega}{\varrho}\frac{1}{\sqrt{\mathcal{P}}}\,(\,\omega p^2\partial_{\tau}+\partial_{\sigma})}\,.
 \end{array}
 \label{ref+}
\end{equation} 
Using (\ref{nice}), (\ref{normal to scri}) it is obvious that ${{\bf t}_{\rm o}={\bf n}=\Omega\,\tilde{\bf n}}$ on $\scri$, i.e. ${\bf t}_{\rm o}$ is a future-oriented timelike normal to $\scri$. Also, 
${\,{\bf q}_{\rm s}={\bf q}_{\rm o}\,}$, ${\,{\bf s}_{\rm s}={\bf s}_{\rm o}\,}$,  and
\begin{equation}
{\bf t}_{\rm s}= \cos^{-1}\!\theta_{\rm s}\,{\bf t}_{\rm o} +  \tan\theta_{\rm s}\,{\bf r}_{\rm o},\qquad
{\bf r}_{\rm s}= \cos^{-1}\!\theta_{\rm s}\,{\bf r}_{\rm o} +  \tan\theta_{\rm s}\,{\bf t}_{\rm o},\label{sp+ref}
\end{equation}
where
\begin{equation}
\sin\theta_{\rm s}= \sqrt{\alpha^2\frac{\,\,\mathcal{P}}{-\mathcal{Q}}},\qquad
\cos\theta_{\rm s}= \sqrt{1-\alpha^2\frac{\,\,\mathcal{P}}{-\mathcal{Q}}}.\label{sincos+}
\end{equation}
(These are explicit relations of those given generally in \cite{KrtousPodolsky:2005}, see equation (21) therein.) Consequently, the \emph{magnitude of the radiative components} of the gravitational and electromagnetic fields are described by the expressions (27), (28) therein, namely
\begin{equation}
|{\Psi }_4^{\rm i}| = \frac{1}{|\eta|}\frac{\frac{3}{2}|{\Psi }_{2\ast}^{\rm s}|}{\cos^2\!\theta_{\rm s}}\,\mathcal{A}(\theta,\phi,\theta_{\rm s})\,,\ \quad
|{\Phi }_2^{\rm i}|^2= \frac{1}{\eta^2}\frac{|{\Phi }_{1\ast}^{\rm s}|^2}{\cos^2\!\theta_{\rm s}}\,\mathcal{A}(\theta,\phi,\theta_{\rm s})\,,\label{asyfield1+}
\end{equation}
where
\begin{equation}
\mathcal{A}(\theta,\phi,\theta_{\rm s})=(\sin\theta+\sin\theta_{\rm s}\cos\phi)^2+\sin^2\theta_{\rm s}\cos^2\!\theta\sin^2\!\phi\,.\label{amplitudseA}
\end{equation}

These are leading components of the fields (with the fall-off of order $\eta^{-1}$) with respect to the interpretation frame parallelly transported along a null geodesic approaching~$\scri$, which has an affine parameter $\eta$. The spherical angles ${\theta,\phi}$ describe the spatial direction on $\scri$ of the null geodesic with respect to the reference tetrad (\ref{ref+}), 
${{\bf q}= \cos\theta\,{\bf q}_{\rm o} +  \sin\theta\,(\cos\phi\,{\bf r}_{\rm o}+\sin\phi\,{\bf s}_{\rm o})}$, and the parameter $\theta_{\rm s}$ is given by (\ref{sincos+}). The coefficient ${\Psi }_{2\ast}^{\rm s}$ is the part of ${\Psi }_{2}$ given by (\ref{weyl1alter}) that is proportional to $\eta^{-3}$, while the coefficient ${\Phi }_{1\ast}^{\rm s}$ is the part of ${\Phi }_{1}$ given by (\ref{Phi1explicit}) that is proportional to $\eta^{-2}$. Since ${\Omega=\epsilon\eta^{-1}\,}$ near $\scri$ (${\epsilon=+1}$ for outgoing null geodesics while ${\epsilon=-1}$ for ingoing null geodesics), see equation (2.19) in \cite{KrtousPodolsky:review}, we immediately conclude that 
\begin{eqnarray}
{\Psi }_{2\ast}^{\rm s}&=&\frac{-\epsilon}{(1-\hbox{i}\,\omega pq)^3}\left[(m+\hbox{i}\,n)+(e^2+g^2)\left(\frac{q-\alpha p}{1+\hbox{i}\,\omega pq}\right)\right],\label{weyl1alterspec}\\
{\Phi }_{1\ast}^{\rm s}&=& \frac{ e(1-\omega^2p^2q^2) - 2g \omega pq
+\hbox{i}\,[\,g(1-\omega^2p^2q^2) + 2e \omega pq\, ]}{2(1+\omega^2p^2q^2)^2}\,.\label{elmag1alterspec}
\end{eqnarray}

Now, it only remains to evaluate the expressions (\ref{sincos+}), (\ref{weyl1alterspec}), and (\ref{elmag1alterspec}) on $\scri$. Using (\ref{skraj}), that is by substituting ${q=-\alpha p}$, all these functions become dependent only on $p$ which specifies various points of scri for such (axially symmetric) spacetimes. Straightforward calculations lead to the following \emph{gravitational} radiation component
\begin{equation}
|{\Psi }_4^{\rm i}| = \frac{1}{|\eta|}\,{\mathcal G}(p)\,\mathcal{A}(\theta,\phi,\theta_{\rm s})\,,\label{asyfield1+final}
\end{equation}
where
\begin{equation}
{\mathcal G}(p) = {\textstyle\frac{3}{2}}\,\sqrt{\mathcal{D}(p)}\,
\frac{\>\alpha^2\mathcal{P}(p)+\frac{\Lambda}{3}(1+\alpha^2\omega^2p^4)\>}{\frac{\Lambda}{3}(1+\alpha^2\omega^2p^4)^{5/2}}\,,\label{calGfinal}
\end{equation}
 with
\begin{equation}
 \hskip-5mm
\mathcal{D}(p)=m^2+n^2-4\alpha(e^2+g^2)\frac{m+n\alpha\omega p^2}{1+\alpha^2\omega^2p^4}\,p
+4\alpha^2\frac{(e^2+g^2)^2}{1+\alpha^2\omega^2p^4}\, p^2\,,\label{Dobecne}
\end{equation}
and $\mathcal{A}(\theta,\phi,\theta_{\rm s})$ is given by (\ref{amplitudseA}) in which, using (\ref{nice}),
\begin{equation}
\sin^2\theta_{\rm s}= \frac{\alpha^2\mathcal{P}(p)}{\alpha^2\mathcal{P}(p)+\frac{\Lambda}{3}(1+\alpha^2\omega^2 p^4)}\,.\label{sincos+final}
\end{equation}
Similarly, for the radiative component of the \emph{electromagnetic} field we obtain
\begin{equation}
|{\Phi }_2^{\rm i}|^2 = \frac{1}{\eta^2}\,{\mathcal E}(p)\,\mathcal{A}(\theta,\phi,\theta_{\rm s})\,,\label{asyfieldEM1+final}
\end{equation}
where
\begin{equation}
{\mathcal E}(p) = {\textstyle\frac{1}{4}}\,(e^2+g^2)\,
\frac{\>\alpha^2\mathcal{P}(p)+\frac{\Lambda}{3}(1+\alpha^2\omega^2p^4)\>}{\frac{\Lambda}{3}(1+\alpha^2\omega^2p^4)^3}\,. \label{calEfinal}
\end{equation}
The function $\mathcal{P}(p)$ has the form (\ref{PQeqnsNEW}) with (\ref{p1})--(\ref{p3}). These expressions give an explicit formula for the radiative components of the gravitational and electromagnetic fields in the complete family of Pleba\'{n}ski--Demia\'{n}ski black hole spacetimes, characterized by the physical parameters $m$, $\alpha$, $a$, $l$, $e$, $g$, $\Lambda$, in the ``de Sitter-like'' case $\Lambda>0$. The amplitude of radiation ${{\mathcal G}(p)}$ will be analyzed and discussed in sections~\ref{sec:amplituderadiation} and~\ref{sec:amplitudevisualPOSITIVE}.

\subsection{The structure of radiation when ${\Lambda<0}$ and ${\,\mathcal{Q}_{\scri} > 0}$}
\label{sec:antideSitterradiationI}

In the case when the cosmological constant $\Lambda$ is negative and the acceleration parameter $\alpha$ is \emph{small}, such that the condition (\ref{znQ>0}) holds everywhere, all regions near the anti-de~Sitter-like (timelike) conformal infinity are stationary and ${\,\mathcal{Q}_{\scri} > 0}$. 

To obtain the corresponding radiation components in these Pleba\'{n}ski--Demia\'{n}ski spacetimes, we will first employ the special null frame ${{\bf k},{\bf l},{\bf m},{\bf \bar{m}}}$ given by (\ref{altervectorPLUS}) for which the function $\mathcal{Q}$ is \emph{positive}. The orthonormal tetrad ${{\bf t}_{\rm s},{\bf q}_{\rm s},{\bf r}_{\rm s},{\bf s}_{\rm s}}$ associated with such principal null directions, introduced by (\ref{orntonullvector}), is thus
\begin{equation}
 \begin{array}{l}
 {\bf t}_{\rm s}={\displaystyle  \frac{\Omega}{\varrho}\frac{1}{\sqrt{\mathcal{Q}}}\,(\,\partial_{\tau}-\omega q^2\partial_{\sigma} )}\,, \\[8pt]
 {\bf q}_{\rm s}={\displaystyle  \frac{\Omega}{\varrho}\sqrt{{\mathcal{Q}}}\,\partial_{q}}\,, \\[8pt]
 {\bf r}_{\rm s}=-{\displaystyle  \frac{\Omega}{\varrho}\sqrt{\mathcal{P}}\,\partial_{p}}\,, \\[8pt]
 {\bf s}_{\rm s}={\displaystyle  \frac{\Omega}{\varrho}\frac{1}{\sqrt{\mathcal{P}}}\,(\,\omega p^2\partial_{\tau}+\partial_{\sigma})}\,.
 \end{array}
 \label{sp+L<0}
\end{equation} 
In the present case, it is convenient to choose the reference orthonormal tetrad
\begin{equation}
 \begin{array}{l}
 {\bf t}_{\rm o}={\displaystyle  \frac{\Omega}{\varrho}\frac{1}{\sqrt{\mathcal{Q}}}\,(\,\partial_{\tau}-\omega q^2\partial_{\sigma} )}\,, \\[8pt]
 {\bf q}_{\rm o}={\displaystyle  \frac{\Omega}{\varrho}\frac{1}{\sqrt{{\mathcal{Q}+\alpha^2\mathcal{P}}}}
\,(\mathcal{Q}\,\partial_{q}+\alpha\mathcal{P}\,\partial_{p})}\,, \\[8pt]
 {\bf r}_{\rm o}=-{\displaystyle  \frac{\Omega}{\varrho}\sqrt{ \frac{\mathcal{P}\mathcal{Q}}{\mathcal{Q}+\alpha^2\mathcal{P}}  }\,(\,\partial_{p}-\alpha \partial_{q})}\,, \\[8pt]
 {\bf s}_{\rm o}={\displaystyle  \frac{\Omega}{\varrho}\frac{1}{\sqrt{\mathcal{P}}}\,(\,\omega p^2\partial_{\tau}+\partial_{\sigma})}\,.
 \end{array}
 \label{ref+L<0}
\end{equation}
Obviously, ${\,{\bf t}_{\rm s}={\bf t}_{\rm o}}$, ${\,{\bf s}_{\rm s}={\bf s}_{\rm o}}$. Using (\ref{normal to scri}), ${{\bf q}_{\rm o}=-{\bf n}\,}$ on $\scri\,$, i.e. ${\bf q}_{\rm o}$ is an \emph{outer} normal to scri (see Fig.~1 and equation (4) in~\cite{KrtousPodolsky:2005}), which corresponds to the choice ${\epsilon_0=+1}$. Since ${\,-{g}_{\alpha\beta}\,{k}^{\alpha}{n}^{\beta}={g}_{\alpha\beta}\,{l}^{\alpha}{n}^{\beta}>0\,}$ in this case, the principal null direction ${\bf k}$ is \emph{outgoing} with respect to $\scri$ whereas the principal null direction ${\bf l}$ is \emph{ingoing}, that is ${\epsilon_1=-\epsilon_2}$, as in relation (29) in~\cite{KrtousPodolsky:2005}. Moreover, 
\begin{equation}
\hskip-14mm
{\bf q}_{\rm s}=  \cosh^{-1}\!\psi_{\rm s}\,{\bf q}_{\rm o} +  \tanh\psi_{\rm s}\,{\bf r}_{\rm o},\qquad
{\bf r}_{\rm s}= -\tanh\psi_{\rm s}\,{\bf q}_{\rm o} +  \cosh^{-1}\!\psi_{\rm s}\,{\bf r}_{\rm o},\label{sp+refLI}
\end{equation}
where
\begin{equation}
\sinh\psi_{\rm s}= \sqrt{\alpha^2\frac{\,\,\mathcal{P}}{\mathcal{Q}}},\qquad
\cosh\psi_{\rm s}= \sqrt{1+\alpha^2\frac{\,\,\mathcal{P}}{\mathcal{Q}}}, \label{sinhcosh+LI}
\end{equation}
which are relations of the form (31) and (10) therein. Thus, the magnitude of the radiative components of the gravitational and electromagnetic fields are given by
\begin{equation}
 \hskip-10mm
|{\Psi }_4^{\rm i}| = \frac{1}{|\eta|}\frac{\frac{3}{2}|{\Psi }_{2\ast}^{\rm s}|}{\cosh^2\!\psi_{\rm s}}\,\mathcal{A}_1(\psi,\phi,\psi_{\rm s})\,,\ \quad
|{\Phi }_2^{\rm i}|^2= \frac{1}{\eta^2}\frac{|{\Phi }_{1\ast}^{\rm s}|^2}{\cosh^2\!\psi_{\rm s}}\,\mathcal{A}_1(\psi,\phi,\psi_{\rm s})\,,\label{asyfield1+next}
\end{equation}
where
\begin{equation}
\mathcal{A}_1(\psi,\phi,\psi_{\rm s})=(\sinh\psi+\epsilon\sinh\psi_{\rm s}\cos\phi)^2+\sinh^2\psi_{\rm s}\cosh^2\!\psi\sin^2\!\phi.\label{amplitudseAanti}
\end{equation}
In these expressions, the pseudospherical parameters $\psi$, $\phi$ characterize the direction on timelike $\scri$ of a given null geodesic with respect to the reference tetrad (\ref{ref+L<0}), namely 
${{\bf t}= \cosh\psi\,{\bf t}_{\rm o} +  \sinh\psi\,(\cos\phi\,{\bf r}_{\rm o}+\sin\phi\,{\bf s}_{\rm o})}$. The parameter $\epsilon$ indicates its orientation: for ${\epsilon=+1}$ the direction is outgoing (pointing outside the spacetime) whereas for ${\epsilon=-1}$ it is an ingoing direction.

The coefficient ${\Psi }_{2\ast}^{\rm s}$ in (\ref{asyfield1+next}) is again given by (\ref{weyl1alterspec}), while ${\Phi }_{1\ast}^{\rm s}$ is now the complex conjugate of (\ref{elmag1alterspec}). Therefore, the radiative components of the gravitational and electromagnetic fields in this case can be written as 
\begin{eqnarray}
|{\Psi }_4^{\rm i}|  &=& \frac{1}{|\eta|}\,|{\mathcal G}(p)|\,\mathcal{A}_1(\psi,\phi,\psi_{\rm s})\,,\label{asyfield1+GRAVLI}\\
|{\Phi }_2^{\rm i}|^2&=& \,\frac{1}{\eta^2}\,|{\mathcal E}(p)|\,\mathcal{A}_1(\psi,\phi,\psi_{\rm s})\,,\label{asyfield1+EMLI}
\end{eqnarray}
where the functions ${\mathcal G}(p)$ and ${\mathcal E}(p)$ have been introduced in (\ref{calGfinal}) and (\ref{calEfinal}). The directional structure 
$\mathcal{A}_1(\psi,\phi,\psi_{\rm s})$ is given by expression (\ref{amplitudseAanti}) in which 
\begin{equation}
\sinh^2\psi_{\rm s}= \frac{\alpha^2\mathcal{P}(p)}{-\frac{\Lambda}{3}(1+\alpha^2\omega^2 p^4)-\alpha^2\mathcal{P}(p)}>0\label{sinhcosh+final}
\end{equation}
determines the orientation of the principal null directions with respect to the conformal infinity $\scri$. These are explicit formulae for the radiative components of the fields in the family of Pleba\'{n}ski--Demia\'{n}ski black hole spacetimes in the ``anti-de~Sitter-like'' case $\Lambda<0$ when the acceleration parameter $\alpha$ is small. In such a situation, ${\mathcal{Q}>0}$ everywhere near $\scri$, on which the principal null direction ${\bf k}$ is outgoing while the principal null direction ${\bf l}$ is ingoing. In the subcase ${\omega=0}$, the results obtained agree with those derived in \cite{PodolskyOrtaggioKrtous:2003}, i.e. for the C-metric with ${\alpha^2<-\frac{\Lambda}{3}}$ which represents a single non-rotating but accelerating black hole in anti-de~Sitter universe.

\subsection{The structure of radiation when ${\Lambda<0}$ and ${\,\mathcal{Q}_{\scri} < 0}$}
\label{sec:antideSitterradiationII}

When the cosmological constant $\Lambda$ is negative but the acceleration parameter $\alpha$ is \emph{large}, there are also regions near the anti-de~Sitter-like (timelike) conformal infinity in which the condition (\ref{znQ<0}) holds. In such regions where ${\,\mathcal{Q}_{\scri} < 0}$, the spacetime is not stationary and the asymptotic structure of radiation is different from that described in previous section~\ref{sec:antideSitterradiationI}.  

To determine the corresponding radiation components we will employ the algebraically special null frame closely related to (\ref{altervectorMINUS}), namely ${{\bf l},{\bf k},-{\bf \bar{m}},-{\bf m}}$. The associated orthonormal tetrad ${{\bf t}_{\rm s},{\bf q}_{\rm s},{\bf r}_{\rm s},{\bf s}_{\rm s}}$ is thus
\begin{equation}
 \begin{array}{l}
{\bf t}_{\rm s}={\displaystyle  \frac{\Omega}{\varrho}\sqrt{{-\mathcal{Q}}}\,\partial_{q}}\,, \\[8pt]
 {\bf q}_{\rm s}={\displaystyle  \frac{\Omega}{\varrho}\frac{1}{\sqrt{-\mathcal{Q}}}\,(\,\partial_{\tau}-\omega q^2\partial_{\sigma} )}\,, \\[8pt]
 {\bf r}_{\rm s}=-{\displaystyle  \frac{\Omega}{\varrho}\sqrt{\mathcal{P}}\,\partial_{p}}\,, \\[8pt]
 {\bf s}_{\rm s}={\displaystyle  \frac{\Omega}{\varrho}\frac{1}{\sqrt{\mathcal{P}}}\,(\,\omega p^2\partial_{\tau}+\partial_{\sigma})}\,,
 \end{array}
 \label{sp-L<0}
\end{equation} 
cf. relations (\ref{orntonullvector}) and (\ref{sp+}). A convenient reference orthonormal tetrad is
\begin{equation}
 \begin{array}{l}
 {\bf t}_{\rm o}=-{\displaystyle  \frac{\Omega}{\varrho}\sqrt{ \frac{-\mathcal{P}\mathcal{Q}}{\mathcal{Q}+\alpha^2\mathcal{P}}  }\,(\,\partial_{p}-\alpha \partial_{q})}\,, \\[8pt]
 {\bf q}_{\rm o}={\displaystyle  \frac{\Omega}{\varrho}\frac{1}{\sqrt{   {\mathcal{Q}}+\alpha^2\mathcal{P}}}
\,(\mathcal{Q}\,\partial_{q}+\alpha\mathcal{P}\,\partial_{p})}\,, \\[8pt]
 {\bf r}_{\rm o}={\displaystyle  \frac{\Omega}{\varrho}\frac{1}{\sqrt{-\mathcal{Q}}}\,(\,\partial_{\tau}-\omega q^2\partial_{\sigma} )}\,, \\[8pt]
 {\bf s}_{\rm o}={\displaystyle  \frac{\Omega}{\varrho}\frac{1}{\sqrt{\mathcal{P}}}\,(\,\omega p^2\partial_{\tau}+\partial_{\sigma})}\,.
 \end{array}
 \label{ref-L<0}
\end{equation}
As in the previous section, using (\ref{normal to scri}) and (\ref{nice}) we obtain ${{\bf q}_{\rm o}=-{\bf n}\,}$ on $\scri\,$, so that ${\epsilon_0=+1}$. However, ${\mathcal{Q}<0}$ which means that ${\bf q}_{\rm o}$ is now the \emph{inner} normal to scri. Also, ${{g}_{\alpha\beta}\,{k}^{\alpha}{n}^{\beta}={g}_{\alpha\beta}\,{l}^{\alpha}{n}^{\beta}<0\,}$ in this case, so that both principal null directions ${\bf k}$ and~${\bf l}$ are \emph{ingoing} with respect to $\scri$ (and ${\epsilon_1=\epsilon_2}$, as in relation (40) in~\cite{KrtousPodolsky:2005}). Moreover, ${\,{\bf q}_{\rm s}={\bf r}_{\rm o}}$, ${\,{\bf s}_{\rm s}={\bf s}_{\rm o}}$, and
\begin{equation}
\hskip-14mm
{\bf t}_{\rm s}=   \coth\psi_{\rm s}\,{\bf t}_{\rm o} + \sinh^{-1}\!\psi_{\rm s}\,{\bf q}_{\rm o},\qquad
{\bf r}_{\rm s}= -\sinh^{-1}\!\psi_{\rm s}\,{\bf t}_{\rm o} -  \coth\psi_{\rm s}\,{\bf q}_{\rm o},\label{sp+refLII}
\end{equation}
where
\begin{equation}
\cosh\psi_{\rm s}= \sqrt{\alpha^2\frac{\,\,\mathcal{P}}{-\mathcal{Q}}}\,,\qquad
\sinh\psi_{\rm s}= \sqrt{\alpha^2\frac{\,\,\mathcal{P}}{-\mathcal{Q}}-1}\,.\label{sinhcosh+LII}
\end{equation}
These are particular exemplifications of relations (42), (10) of~\cite{KrtousPodolsky:2005}. Consequently, the radiative components of the fields are given by expressions
\begin{equation}
 \hskip-10mm
|{\Psi }_4^{\rm i}| = \frac{1}{|\eta|}\frac{\frac{3}{2}|{\Psi }_{2\ast}^{\rm s}|}{\sinh^2\!\psi_{\rm s}}\,\mathcal{A}_2(\psi,\phi,\psi_{\rm s})\,,\ \quad
|{\Phi }_2^{\rm i}|^2= \frac{1}{\eta^2}\frac{|{\Phi }_{1\ast}^{\rm s}|^2}{\sinh^2\!\psi_{\rm s}}\,\mathcal{A}_2(\psi,\phi,\psi_{\rm s})\,,\label{asyfield1+II}
\end{equation}
where
\begin{equation}
\mathcal{A}_2(\psi,\phi,\psi_{\rm s})=(\cosh\psi_{\rm s}+\epsilon\cosh\psi)^2+\sinh^2\psi_{\rm s}\sinh^2\!\psi\sin^2\!\phi\,.\label{amplitudseAantiII}
\end{equation}
Again, the pseudospherical parameters $\psi$, $\phi$ characterize the direction on timelike $\scri$ of a given null geodesic with respect to the reference tetrad (\ref{ref-L<0}), ${\epsilon=\pm1}$ denotes its orientation, the coefficient ${\Psi }_{2\ast}^{\rm s}$ is given by (\ref{weyl1alterspec}), while ${\Phi }_{1\ast}^{\rm s}$ is given by (\ref{elmag1alterspec}) multiplied by the overall factor $(-1)$. 

The radiative components of gravitational and electromagnetic fields in this case are thus
\begin{eqnarray}
|{\Psi }_4^{\rm i}|  &=& \frac{1}{|\eta|}\,|{\mathcal G}(p)|\,\mathcal{A}_2(\psi,\phi,\psi_{\rm s})\,,\label{asyfield1+GRAVLII}\\
|{\Phi }_2^{\rm i}|^2&=& \,\frac{1}{\eta^2}\,|{\mathcal E}(p)|\,\mathcal{A}_2(\psi,\phi,\psi_{\rm s})\,,\label{asyfield1+EMLII}
\end{eqnarray}
where the functions ${\mathcal G}(p)$ and ${\mathcal E}(p)$, now with an opposite sign, have been defined in (\ref{calGfinal}), (\ref{calEfinal}). The directional structure 
$\mathcal{A}_2(\psi,\phi,\psi_{\rm s})$ is now given by expression (\ref{amplitudseAantiII}) in which, using (\ref{sinhcosh+LII}) and (\ref{nice}), 
\begin{eqnarray}
&&\cosh^2\psi_{\rm s}= \frac{\alpha^2\mathcal{P}(p)}{\alpha^2\mathcal{P}(p)+\frac{\Lambda}{3}(1+\alpha^2\omega^2 p^4)}\,,\nonumber\\
&&\sinh^2\psi_{\rm s}= \frac{-\frac{\Lambda}{3}(1+\alpha^2\omega^2 p^4)}{\alpha^2\mathcal{P}(p)+\frac{\Lambda}{3}(1+\alpha^2\omega^2 p^4)}\,.\label{sinhcosh+finalII}\end{eqnarray}
These are explicit formulae for the radiative components of the fields in the family of Pleba\'{n}ski--Demia\'{n}ski black hole spacetimes in the ``anti-de~Sitter-like'' case $\Lambda<0$ when the acceleration parameter $\alpha$ is large and ${\mathcal{Q}<0}$. More precisely, for large $\alpha$ there are both stationary (${\mathcal{Q}>0}$) and non-stationary (${\mathcal{Q}<0}$) regions near the conformal infinity $\scri$. In the present case ${\mathcal{Q}<0}$, so that both the principal null directions ${\bf k}$ and ${\bf l}$ are ingoing on $\scri$. Note that for ${\omega=0}$, the above results (\ref{amplitudseAantiII})--(\ref{sinhcosh+finalII}) reduce to those derived in \cite{PodolskyOrtaggioKrtous:2003}, i.e. for the C-metric with ${\alpha^2>-\frac{\Lambda}{3}}$ which describes a pair of non-rotating, uniformly accelerating black holes in anti-de~Sitter universe.

\section{Discussion of radiative properties of the Pleba\'nski--Demia\'nski black hole spacetimes}
\label{discussion}
In the preceding section~\ref{radiation} we have derived \emph{explicit} forms of radiative components of gravitational and electromagnetic fields for a completely general family of Pleba\'nski--Demia\'nski black hole spacetimes with a cosmological constant and a twist. The asymptotic structure of such fields has been described by their specific components evaluated in the interpretation frame that is parallelly transported along null geodesics approaching~$\scri$. We have denoted them as ${|{\Psi }_4^{\rm i}|}$ and ${|{\Phi }_2^{\rm i}|}$, respectively.

When the cosmological constant is positive (${\Lambda>0}$) the conformal infinity $\scri$ has a de Sitter-like character (i.e., its normal is everywhere timelike) and the asymptotic components of the fields are given by expressions (\ref{asyfield1+final}), (\ref{asyfieldEM1+final}). In the anti-de Sitter-like case (${\Lambda<0}$) the normal to scri is spacelike, and it is necessary to distinguish two cases. For a small acceleration $\alpha$ of the black holes, more precisely in the points of~$\scri$ characterized by the coordinate $p$ where ${\alpha^2(\mathcal{P}(p)+\frac{\Lambda}{3}\omega^2 p^4) <-\frac{\Lambda}{3}}\,$, the asymptotic fields are given by (\ref{asyfield1+GRAVLI}), (\ref{asyfield1+EMLI}), while for a large acceleration there are also regions on $\scri$ where ${\alpha^2(\mathcal{P}(p)+\frac{\Lambda}{3}\omega^2 p^4) >-\frac{\Lambda}{3}\,}$, in which the  fields are given by (\ref{asyfield1+GRAVLII}), (\ref{asyfield1+EMLII}). 

All these radiative components asymptotically decay as ${\,\eta^{-1}}$, where $\eta$ is an affine parameter along the null geodesic (and ${\,|\eta|\to\infty}$ as the geodesic approaches $\scri$\,).

\subsection{The directional structure of radiation on $\scri$}
\label{sec:directstucture}
When ${\Lambda>0}$, the directional structure of radiation at any point at conformal infinity is fully determined by the function ${\mathcal{A}(\theta,\phi,\theta_{\rm s})}$ given in (\ref{amplitudseA}). In this case, the spherical angles ${\theta,\phi}$ describe the spatial direction ${{\bf q}= \cos\theta\,{\bf q}_{\rm o} +  \sin\theta\,(\cos\phi\,{\bf r}_{\rm o}+\sin\phi\,{\bf s}_{\rm o})}$ on~$\scri$ of a given null geodesic with respect to the reference orthonormal tetrad (\ref{ref+}). The parameter $\theta_{\rm s}$ is explicitly determined by (\ref{sincos+final}) --- let us recall that the variable~$p$ therein determines the specific point on $\scri$ where the null geodesic ``terminates''. Such a directional structure of radiation (which is visualized on Fig.~3 of~\cite{KrtousPodolsky:2005}), is typical for all type~D spacetimes that admit a spacelike conformal infinity. 

When ${\Lambda<0}$ and acceleration $\alpha$ is small, the directional structure of radiation on a timelike conformal infinity is described by the function ${\mathcal{A}_1(\psi,\phi,\psi_{\rm s})}$ introduced in (\ref{amplitudseAanti}). The pseudospherical parameters $\psi$, $\phi$ characterize the direction on $\scri$ of a given null geodesic with respect to the reference tetrad~(\ref{ref+L<0}), ${{\bf t}= \cosh\psi\,{\bf t}_{\rm o} +  \sinh\psi\,(\cos\phi\,{\bf r}_{\rm o}+\sin\phi\,{\bf s}_{\rm o})}$, and ${\psi_{\rm s}}$ given by (\ref{sinhcosh+final}) determines the orientation of the principal null directions with respect to $\scri$ (here ${\bf k}$ is outgoing while ${\bf l}$ is ingoing since ${\mathcal{Q}>0}$ everywhere near $\scri$\,). However, when the acceleration $\alpha$ is large, there are also non-stationary regions near timelike conformal infinity where ${\mathcal{Q}<0}$ (and both ${\bf k}$ and ${\bf l}$ are outgoing), in which case the asymptotic directional structure of radiation at a given point on $\scri$ is described by ${\mathcal{A}_2(\psi,\phi,\psi_{\rm s})}$. This is given by (\ref{amplitudseAantiII}), with the parameter ${\psi_{\rm s}}$ determined by (\ref{sinhcosh+finalII}). Again, these directional structures (shown on Fig.~5 of~\cite{KrtousPodolsky:2005}) are characteristic for type~D spacetimes with a timelike $\scri$.

\subsection{The amplitude of radiation on $\scri$}
\label{sec:amplituderadiation}

The main result of the present work is the derivation of explicit asymptotic radiative component of the gravitational field in the complete family of Pleba\'nski--Demia\'nski black hole spacetimes with a possible twist and a cosmological constant $\Lambda$. This is uniquely determined by the amplitude ${{\mathcal G}(p)}$ given by (\ref{calGfinal}). The analogous radiative component of the electromagnetic field is given by the amplitude ${{\mathcal E}(p)}$, see (\ref{calEfinal}). Interestingly, the functions ${|{\mathcal G}|}$ and ${|{\mathcal E}|}$ are \emph{the same} for all three possibilities discussed above, namely ${\Lambda>0}$ and ${\Lambda<0}$ with either ${\mathcal{Q}>0}$ or ${\mathcal{Q}<0}$.

In the remaining part of this contribution we will thus concentrate on description and analysis of such amplitudes. In particular, we will investigate and visualize these functions which uniquely express the dependence of the magnitude of gravitational and electromagnetic radiation on the parameters that characterize various physical properties of the black holes, namely their mass $m$, electric and magnetic charges $e$ and $g$, NUT parameter $l$, rotational parameter $a$, and acceleration $\alpha$. Within the framework of rigorous Einstein theory, we may thus study specific properties of radiation generated by localized sources of the black hole type, as described by the metric (\ref{PleDemMetricNEW}), (\ref{PQeqnsNEW}).

The \emph{amplitude of gravitational radiation} ${{\mathcal G}(p)}$ is given by expression (\ref{calGfinal}) in which the specific function $\mathcal{D}(p)$ takes the form (\ref{Dobecne}). The parameter $p$, together with the angular coordinate ${\sigma=-(\omega/a)\,\phi}$, where ${\phi\in[0,2\pi)}$, labels all possible points on the (axially symmetric) conformal infinity $\scri$. As explained in section~\ref{genPDfamily}, the coordinate $p$ must be restricted to a particular range between appropriate roots ${\mathcal{P}=0}$ which correspond to poles of the black holes. This is automatically achieved by assuming the parameterization ${\,p=(l/\omega)+(a/\omega)\cos\theta\,}$ with ${\,\theta\in[0,\pi]}$, see transformation (\ref{trans1A}). Simultaneously, it is necessary to express the (unphysical) coefficients $\epsilon$, $n$ and $k$ in (\ref{PQeqnsNEW}) using the expressions (\ref{p1})--(\ref{p3}).

The resulting general expression for ${{\mathcal G}(p(\theta))}$ is involved. For simplicity, we will thus restrict our attention only to the  case, in which the \emph{NUT parameter $l$ vanishes}. Setting ${l=0}$, the black hole spacetimes are free from torsion singularities, regions with closed timelike curves are absent, and the spacetimes are physically more realistic. With this assumption, the twist parameter $\,\omega$ can be chosen to be equal to~$\,a$, so that 
\begin{equation}
p=\cos\theta\,.
\end{equation}
The complete range ${\,\theta\in[0,\pi]}$ thus corresponds to ${\,p\in[-1,+1]}$. The values ${p=\pm1}$ represent \emph{poles} of the black hole spacetimes, while the special value ${p=0}$ describes the \emph{equatorial plane} ${\theta=\frac{\pi}{2}}$. 

In such a case, expressions (\ref{p1})--(\ref{p3}) simplify considerably to
\begin{equation}
  k=1\,,\quad
  n=-\alpha a m,\quad
\epsilon=1-\alpha^2(e^2+g^2)-\big(\alpha^2+{\textstyle\frac{\Lambda}{3}}\big)a^2,\label{pfin}
\end{equation}
the function $\mathcal{P}(p)$, given by (\ref{PQeqnsNEW}), becomes
\begin{equation}
\mathcal{P}(p)=(1-p^2)\Big( 1 - 2\alpha m\, p + \big[\,\alpha^2(e^2+g^2)+\big(\alpha^2+{\textstyle\frac{\Lambda}{3}}\big)a^2\,\big]\, p^2\Big), 
 \label{PQeqnsNEWfin}
\end{equation}
and the function $\mathcal{D}(p)$, introduced in (\ref{Dobecne}), takes the form
\begin{equation}
  \hskip-14mm
\mathcal{D}(p)=m^2(1+\alpha^2 a^2)-4\alpha m(e^2+g^2)\frac{1- \alpha^2a^2 p^2}{1+\alpha^2a^2p^4}\,p
+4\alpha^2\frac{(e^2+g^2)^2}{1+\alpha^2a^2p^4}\, p^2.
\label{Dobecnedis}
\end{equation}
The amplitude (\ref{calGfinal}) is therefore
\begin{equation}
{\mathcal G}(p) = {\textstyle\frac{3}{2}}\,\sqrt{\mathcal{D}(p)}\,
\frac{\>\alpha^2\mathcal{P}(p)+\frac{\Lambda}{3}(1+\alpha^2a^2p^4)\>}{\frac{\Lambda}{3}(1+\alpha^2a^2p^4)^{5/2}}\,.\label{calGfinaldis}
\end{equation}
The above expressions (\ref{PQeqnsNEWfin})--(\ref{calGfinaldis}) explicitly describe the amplitude of gravitational radiation in the spacetimes with a cosmological constant $\Lambda$ which represent uniformly accelerating and rotating charged black holes, that are characterized by the parameters $\alpha$, $a$, $e$, $g$ and $m$, respectively. In particular, in the equatorial plane given by ${p=0}$, the amplitude simplifies considerably to 
\begin{equation}
{\mathcal G}(0) = {\textstyle\frac{3}{2}}\,m\,\sqrt{1+\alpha^2 a^2}\,
(\,1+{\textstyle\frac{3}{\Lambda}}\alpha^2)\,.\label{calGfinalequator}
\end{equation}

In the case of \emph{uncharged} black holes (${e=0=g}$), the amplitude of gravitation (\ref{calGfinaldis}) reduces to   
\begin{equation}
{\mathcal G}(p) = {\textstyle\frac{3}{2}}\,m\,\sqrt{1+\alpha^2 a^2}\,
\frac{\>\alpha^2\mathcal{P}(p)+\frac{\Lambda}{3}(1+\alpha^2a^2p^4)\>}{\frac{\Lambda}{3}(1+\alpha^2a^2p^4)^{5/2}}\,,\label{calGfinalunchp}
\end{equation}
where
\begin{equation}
\mathcal{P}(p)=(1-p^2)\Big( 1 - 2\alpha m\, p + \big(\alpha^2+{\textstyle\frac{\Lambda}{3}}\big)a^2\, p^2\Big)\,. 
 \label{PQeqnsNEWunch}
\end{equation}
These expressions describe the amplitude of gravitational radiation generated by Kerr-like black holes (of mass $m$ and rotation parameter $a$) which uniformly accelerate (as determined by the acceleration parameter $\alpha$) in the de~Sitter or anti-de~Sitter universe with a cosmological constant $\Lambda$.  

For a \emph{vanishing rotation} of the black holes, ${a=0}$, this further simplifies to 
\begin{equation}
{\mathcal G}(p) = {\textstyle\frac{3}{2}}\,m
\Big(1+{\textstyle\frac{3}{\Lambda}}\alpha^2(1-p^2)(1 - 2\alpha mp)\Big)\,,\label{calGfinalunchnonrot}
\end{equation}
which corresponds to the uncharged C-metric, while for a \emph{vanishing acceleration}, ${\alpha=0}$, we immediately obtain
\begin{eqnarray}
{\mathcal G}(p) &=& {\textstyle\frac{3}{2}}\,m\,,\qquad\hbox{and}\nonumber\\
\quad\theta_{\rm s}&=&0\quad\hbox{when}\quad \Lambda>0,\\
\quad\psi_{\rm s}  &=&0\quad\hbox{when}\quad \Lambda<0,\nonumber
\label{calGfinalunchnonaccel}
\end{eqnarray}
which corresponds to the Kerr--(anti-)de~Sitter black holes. In this case, the amplitude of gravitational radiation ${\mathcal G}$ is constant (proportional to the mass parameter $m$ of the black hole), independent of the rotation parameter $a$ and the specific value of the cosmological constant $\Lambda$. Moreover, it is also \emph{independent of the position $p$ on scri} $\scri$. The directional structure of radiation on $\scri$ is given by ${\mathcal{A}(\theta)=\,\sin^2\theta}$ or ${\mathcal{A}_1(\psi)=\,\sinh^2\psi}$, if $\Lambda$ is positive or negative, respectively.

\subsection{Visualisation of the results for ${\Lambda>0}$}
\label{sec:amplitudevisualPOSITIVE}

\begin{figure}[htp]
\begin{center}
\includegraphics[scale=0.53]{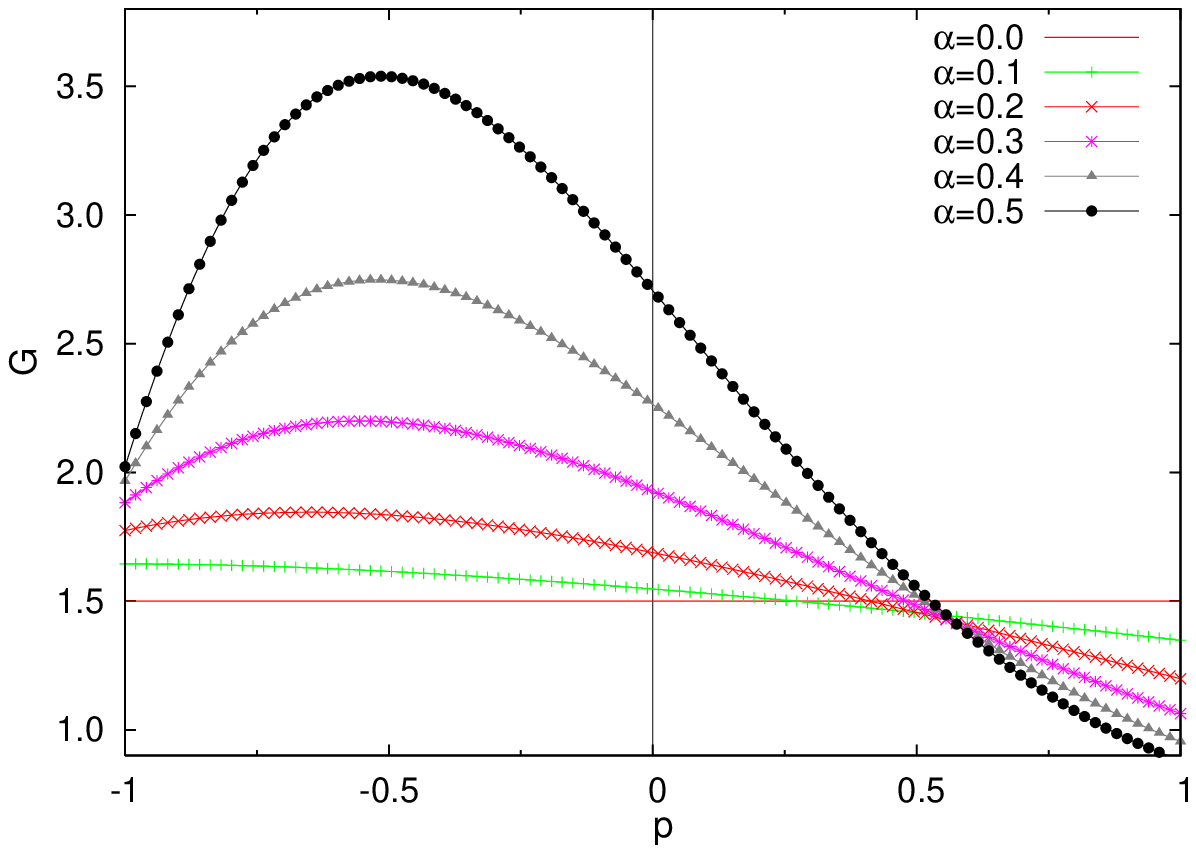}
\end{center}
\caption{\label{fig:graf1}%
The amplitude of gravitational radiation $\mathcal{G}(p)$ on a spacelike $\mathcal{I}$, in which case ${\mathcal{Q}_{\mathcal{I}}<0}$ for any point $p$ of $\scri$. As a typical example, it is assumed that ${\Lambda=1}$, ${m=1}$, ${a=0.5}$, ${e=0.5=g}$, while the acceleration of black holes is taken in the interval ${\alpha\in[0,0.5]}$. The horizontal line at ${\mathcal{G}=1.5}$ represents the uniform amplitude for non-accelerated Kerr--Newman--de-Sitter spacetime with ${\alpha=0}$.
}
\end{figure}

In this case, for a \emph{generic} choice of the parameters of rotating charged black holes which accelerate in de Sitter universe, the function ${\mathcal G}(p)$ determined by the expression (\ref{calGfinaldis}) is plotted in figure~\ref{fig:graf1}. As a typical example, the values are taken as ${\Lambda=1}$, ${m=1}$, ${a=0.5}$, ${e=0.5=g}$, and ${\alpha=0, 0.1, 0.2, 0.3, 0.4, 0.5}$. It can be seen that the amplitude of gravitation radiation has a single maximum around the value ${p\approx -0.5}$ (i.e. in the ``southern hemisphere'' of $\scri\,$) and decreases monotonously to a minimal value at the north pole ${p=1}$. The greater the acceleration $\alpha$, the bigger the generated radiation.  

\begin{figure}[htp]
\begin{center}
\includegraphics[scale=0.53]{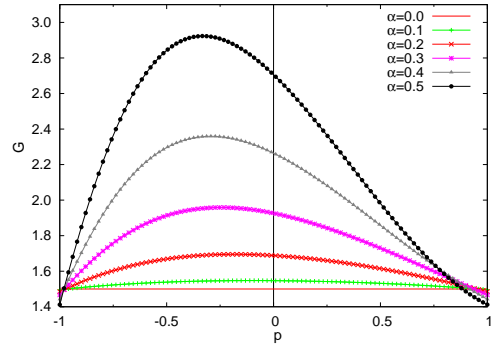}
\end{center}
\caption{\label{fig:graf2}%
The amplitude $\mathcal{G}(p)$ on a spacelike $\mathcal{I}$ for ${\Lambda=1}$, ${m=1}$, ${a=0.5}$, ${e=0=g}$, and ${\alpha\in[0,0.5]}$. This describes gravitational radiation of uncharged rotating black holes which accelerate in de-Sitter spacetime.
}
\end{figure}
\begin{figure}[htp]
\begin{center}
\includegraphics[scale=0.53]{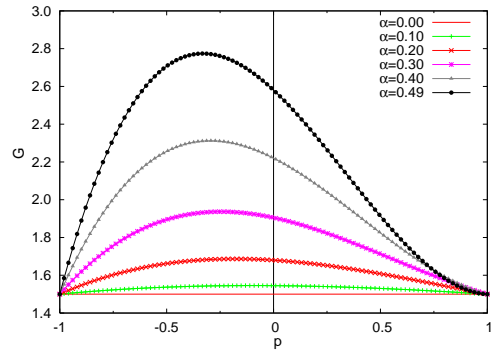}
\end{center}
\caption{\label{fig:graf4}%
The amplitude $\mathcal{G}(p)$ on a spacelike $\mathcal{I}$ for ${\Lambda=1}$, ${m=1}$, ${a=0}$, ${e=0=g}$, and ${\alpha\in[0,0.49]}$. This represents radiation of non-rotating uncharged black holes which accelerate in de-Sitter spacetime, described by the C-metric.
}
\end{figure}

\begin{figure}[htp]
\begin{center}
\includegraphics[scale=0.53]{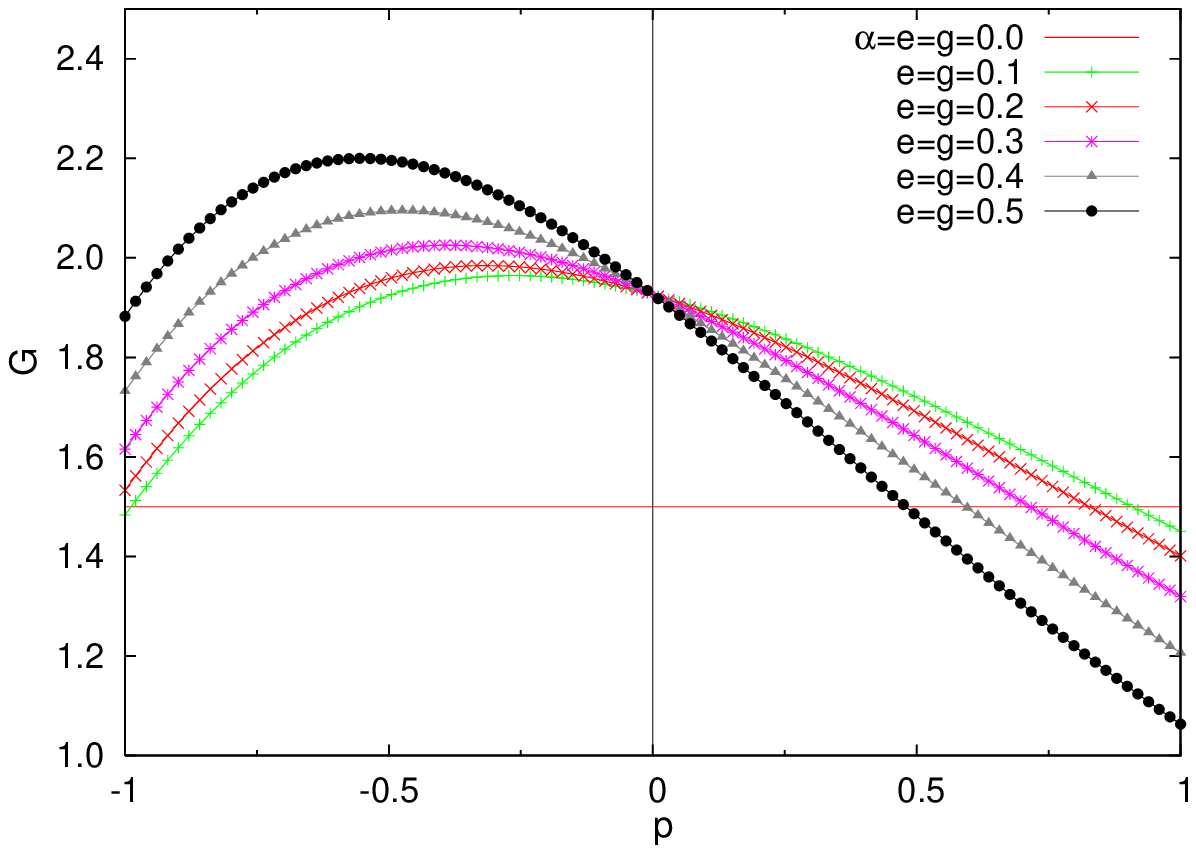}
\end{center}
\caption{\label{fig:graf7}%
The amplitude $\mathcal{G}(p)$ on a spacelike $\mathcal{I}$ for ${\Lambda=1}$, ${m=1}$, ${a=0.5}$, ${\alpha=0.3}$, and  ${e=g\in[0,0.5]}$. This shows the dependence of gravitational radiation, generated by accelerating rotating black holes, on their charges.
}
\end{figure}

\begin{figure}[htp]
\begin{center}
\includegraphics[scale=0.53]{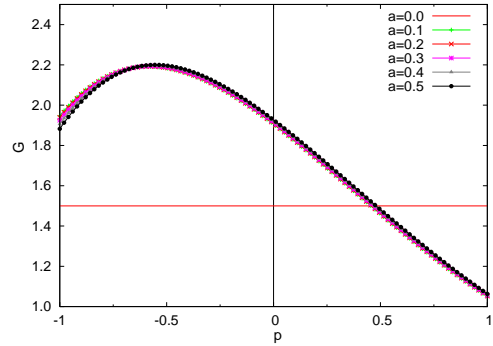}
\end{center}
\caption{\label{fig:graf5}%
The amplitude $\mathcal{G}(p)$ on a spacelike $\mathcal{I}$ for ${\Lambda=1}$, ${m=1}$, ${e=0.5=g}$, ${\alpha=0.3}$, and  ${a\in[0,0.5]}$. The dependence of gravitational radiation, generated by accelerating charged black holes, on their rotation $a$ is very weak.
}
\end{figure}

Similar results are obtained for \emph{uncharged} rotating black holes (${e=0=g}$), see figure~\ref{fig:graf2} plotting (\ref{calGfinalunchp}), and \emph{non-rotating} black holes (${a=0}$), see figure~\ref{fig:graf4} for the case of an uncharged C-metric. While the dependence on the acceleration $\alpha$ is quite strong, the dependence on the charge parameters $e$, $g$ is weaker, see figure~\ref{fig:graf7}, and the dependence on the rotation parameter $a$ is much weaker, see figure~\ref{fig:graf5}.


\subsection{Visualisation of the results for ${\Lambda<0}$ and ${\,\mathcal{Q}_{\scri} > 0}$}
\label{sec:amplitudevisualNEGATIVE+}

\begin{figure}[htp]
\begin{center}
\includegraphics[scale=0.53]{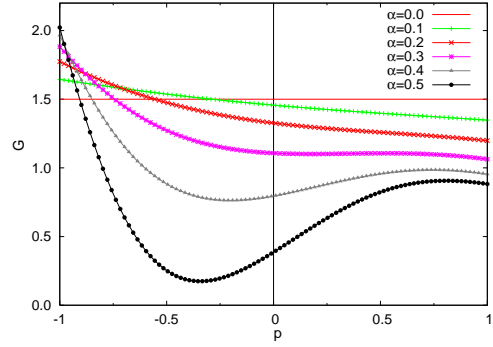}
\end{center}
\caption{\label{fig:graf9}%
The amplitude of gravitational radiation $\mathcal{G}(p)$ on a timelike $\mathcal{I}$ in the case when ${\mathcal{Q}_{\mathcal{I}}>0}$. As a typical example, it is assumed that ${\Lambda=-1}$, ${m=1}$, ${a=0.5}$, ${e=0.5=g}$, while the acceleration of black holes is taken in the interval ${\alpha\in[0,0.5]}$. The horizontal line at ${\mathcal{G}=1.5}$ represents the uniform amplitude for non-accelerated Kerr--Newman--anti-de-Sitter spacetime with ${\alpha=0}$.
}
\end{figure}

When the cosmological constant $\Lambda$ is negative and the acceleration parameter $\alpha$ is small enough, such that the condition (\ref{znQ>0}) holds everywhere on $\scri$, all points $p$ on the anti-de~Sitter-like (timelike) conformal infinity are in the stationary region since ${\,\mathcal{Q}_{\scri} > 0}$. For a typical choice of the parameters of accelerating rotating charged black holes the function ${\mathcal G}(p)$  is plotted in figure~\ref{fig:graf9}. As an example, the values are taken as ${\Lambda=-1}$, ${m=1}$, ${a=0.5}$, ${e=0.5=g}$, and ${\alpha=0, 0.1, 0.2, 0.3, 0.4, 0.5}$. It can be seen that for very small values of $\alpha$, the amplitude of gravitation radiation decreases uniformly from the south pole (${p=-1}$) to the north pole (${p=+1}$), while for larger accelerations there is a significant local minimum in the ``southern hemisphere''.  

\begin{figure}[htp]
\begin{center}
\includegraphics[scale=0.53]{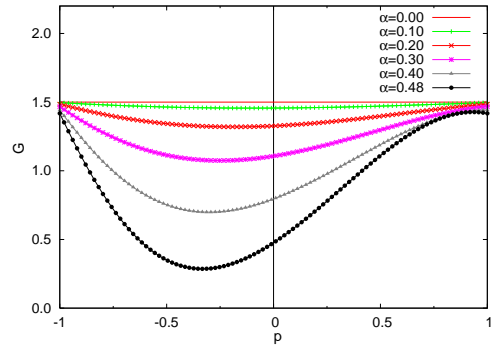}
\end{center}
\caption{\label{fig:graf10}%
The amplitude $\mathcal{G}(p)$ on a timelike $\mathcal{I}$ and ${\mathcal{Q}_{\mathcal{I}}>0}$, for ${\Lambda=-1}$, ${m=1}$, ${a=0.5}$, ${e=0=g}$, and ${\alpha\in[0,0.48]}$. This describes gravitational radiation of uncharged rotating black hole which accelerate in anti-de-Sitter spacetime.
}
\end{figure}
\begin{figure}[htp]
\begin{center}
\includegraphics[scale=0.53]{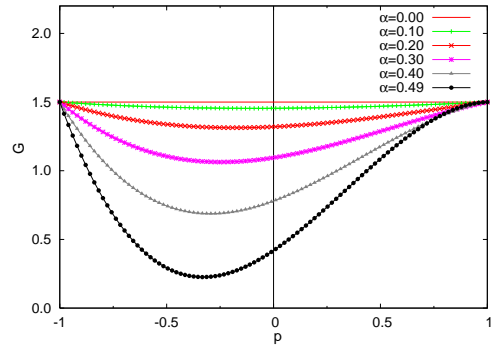}
\end{center}
\caption{\label{fig:graf12}%
The amplitude $\mathcal{G}(p)$ on a timelike $\mathcal{I}$ and ${\mathcal{Q}_{\mathcal{I}}>0}$, for ${\Lambda=-1}$, ${m=1}$, ${a=0}$, ${e=0=g}$, ${\alpha\in[0,0.49]}$. This describes radiation of non-rotating uncharged black holes which accelerate in anti-de-Sitter spacetime (given by the C-metric).
}
\end{figure}

\begin{figure}[htp]
\begin{center}
\includegraphics[scale=0.53]{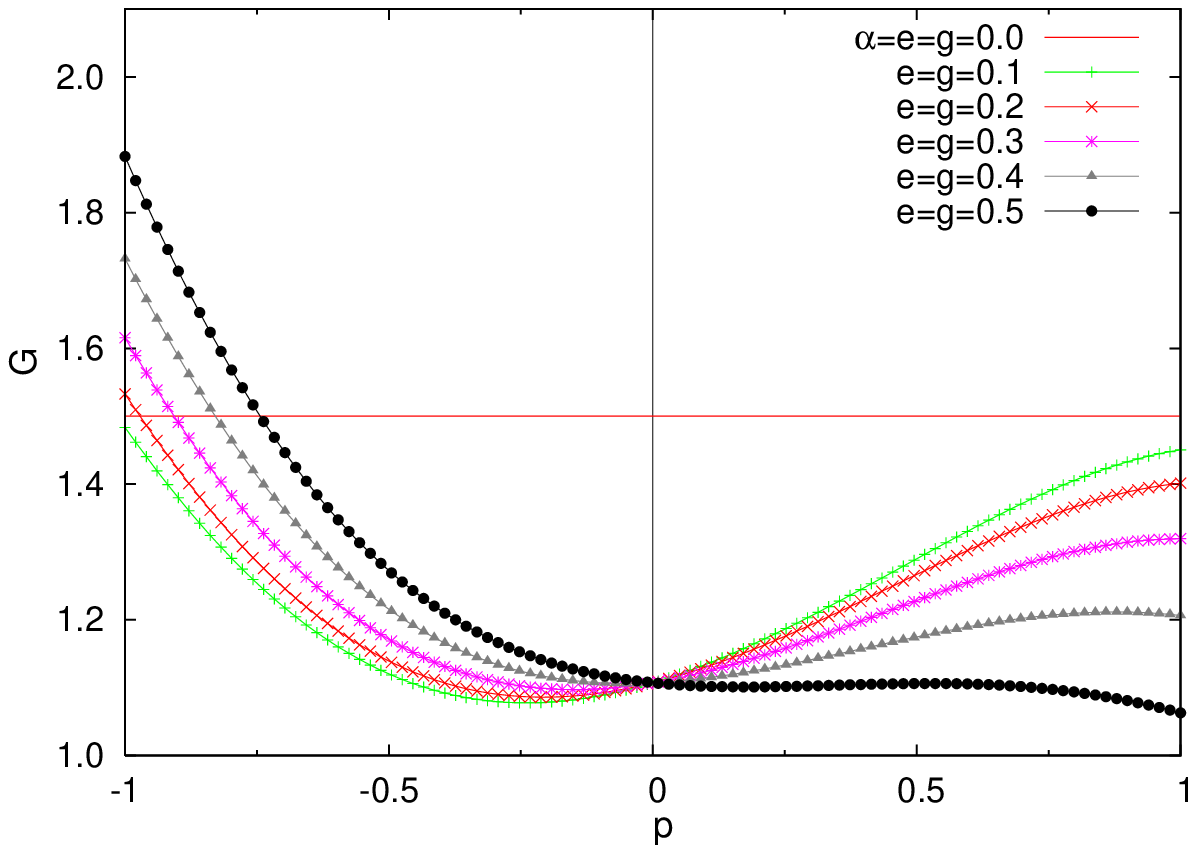}
\end{center}
\caption{\label{fig:graf15}%
The amplitude $\mathcal{G}(p)$ on a timelike $\mathcal{I}$ and ${\mathcal{Q}_{\mathcal{I}}>0}$, for ${\Lambda=-1}$, ${m=1}$, ${a=0.5}$, ${\alpha=0.3}$, and  ${e=g\in[0,0.5]}$ which shows the dependence of gravitational radiation, generated by accelerating rotating black holes, on their charges.
}
\end{figure}

\begin{figure}[htp]
\begin{center}
\includegraphics[scale=0.53]{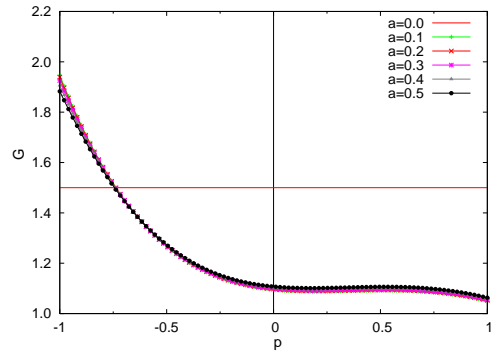}
\end{center}
\caption{\label{fig:graf13}%
The amplitude $\mathcal{G}(p)$ on a timelike $\mathcal{I}$ and ${\mathcal{Q}_{\mathcal{I}}>0}$, for ${\Lambda=-1}$, ${m=1}$, ${e=0.5=g}$, ${\alpha=0.3}$, and  ${a\in[0,0.5]}$. This shows that the dependence of gravitational radiation on rotation of these black holes is very weak.
}
\end{figure}

Analogous results are obtained for \emph{uncharged} rotating black holes (${e=0=g}$), see figure~\ref{fig:graf10}, and \emph{non-rotating} black holes (${a=0}$), see figure~\ref{fig:graf12} which corresponds to the case of an uncharged C-metric. As in the previous case ${\Lambda>0}$, the dependence on the acceleration $\alpha$ is generally strong, the dependence on the charges $e$, $g$ is weaker, see figure~\ref{fig:graf15}, and the dependence on the rotation $a$ is considerably weaker, see figure~\ref{fig:graf13}.


\subsection{Visualisation of the results for ${\Lambda<0}$ and ${\,\mathcal{Q}_{\scri} < 0}$}
\label{sec:amplitudevisualNEGATIVE+-}

When the cosmological constant $\Lambda$ is negative but the acceleration parameter $\alpha$ is large, the condition (\ref{znQ>0}) for ${\,\mathcal{Q}_{\scri} > 0}$ is \emph{not satisfied everywhere} on the anti-de~Sitter-like (timelike) conformal infinity $\scri$. In the regions (i.e. for appropriate ranges of $p$) where ${\,\mathcal{Q}_{\scri} > 0}$, the amplitude of gravitational radiation $\mathcal{G}(p)$ still behaves qualitatively in the same way as described in section~\ref{sec:amplitudevisualNEGATIVE+}. However, now there are also non-stationary regions of $\scri\,$ in which the condition (\ref{znQ<0}) holds so that ${\,\mathcal{Q}_{\scri} < 0}$. (For more details about the structure of the anti-de~Sitter $\scri\,$ for large values of $\alpha$ see \cite{PodolskyOrtaggioKrtous:2003,Krtous:2005}.)

\begin{figure}[htp]
\begin{center}
\includegraphics[scale=0.53]{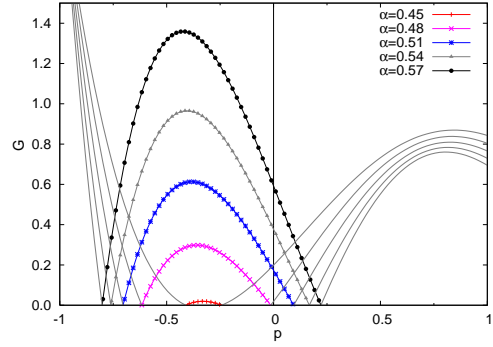}
\end{center}
\caption{\label{fig:graf17}%
The amplitude of gravitational radiation $|\mathcal{G}(p)|$ on a timelike $\mathcal{I}$ in the regions where ${\mathcal{Q}_{\mathcal{I}}<0}$. It is assumed that ${\Lambda=-0.7}$, ${m=1}$, ${a=0.2}$, ${e=0.5=g}$, while the acceleration of the black holes is taken in the interval ${\alpha\in[0.45,0.57]}$. For completeness, we also plot $|\mathcal{G}(p)|$ in the stationary regions where ${\mathcal{Q}_{\mathcal{I}}>0}$ by the corresponding grey lines.
}
\end{figure}

\begin{figure}[htp]
\begin{center}
\includegraphics[scale=0.53]{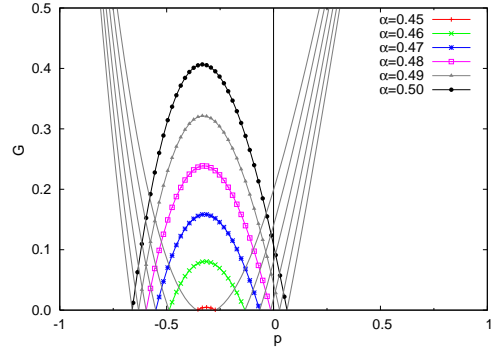}
\end{center}
\caption{\label{fig:graf18}%
The amplitude $|\mathcal{G}(p)|$ on a part of timelike $\mathcal{I}$ where ${\mathcal{Q}_{\mathcal{I}}<0}$, for ${\Lambda=-0.7}$, ${m=1}$, ${a=0.2}$, ${e=0=g}$, and ${\alpha\in[0.45,0.50]}$. It describes radiation of uncharged rotating black holes accelerating in anti-de-Sitter spacetime.
}
\end{figure}

\begin{figure}[htp]
\begin{center}
\includegraphics[scale=0.53]{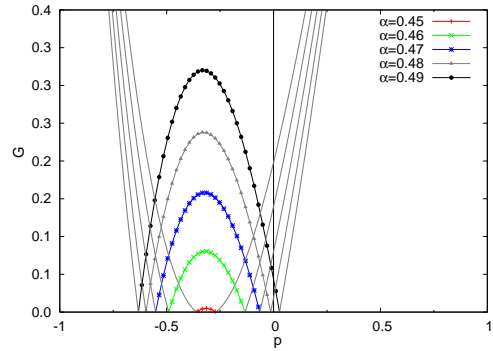}
\end{center}
\caption{\label{fig:graf20}%
The amplitude $|\mathcal{G}(p)|$ on a part of timelike $\mathcal{I}$ where ${\mathcal{Q}_{\mathcal{I}}<0}$, for ${\Lambda=-0.7}$, ${m=1}$, ${a=0}$, ${e=0=g}$, ${\alpha\in[0.45,0.49]}$, describing radiation of non-rotating uncharged black holes accelerating in anti-de-Sitter space (the C-metric).
}
\end{figure}

For a typical such case ${\Lambda=-0.7}$, ${m=1}$, ${a=0.2}$, ${e=0.5=g}$, and ${\alpha=0.45}$, ${0.48, 0.51, 0.54, 0.57}$, the function $|{\mathcal G}(p)|$ is plotted in figure~\ref{fig:graf17}. Admissible ranges of the coordinate $p$ and of the acceleration parameter $\alpha$ result from the necessity to satisfy both the conditions ${\,\mathcal{P} > 0}$ and ${\,\mathcal{Q}_{\scri} < 0}$. There is a single maximum localized on the southern hemisphere, approximately in the centre of the corresponding interval of $p$. It can be seen from figures~\ref{fig:graf18} and~\ref{fig:graf20} that for uncharged (${e=0=g}$) and non-rotating (${a=0}$) black holes the behaviour of the function $|{\mathcal G}(p)|$ is basically the same.


\begin{figure}[htp]
\begin{center}
\includegraphics[scale=0.53]{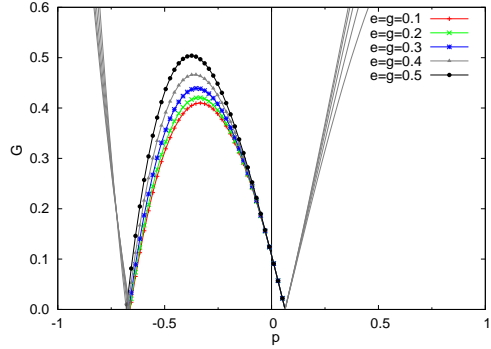}
\end{center}
\caption{\label{fig:graf23}%
The amplitude of gravitational radiation $|\mathcal{G}(p)|$ on a part of timelike $\mathcal{I}$ where ${\mathcal{Q}_{\mathcal{I}}<0}$, for ${\Lambda=-0.7}$, ${m=1}$, ${a=0.2}$, ${\alpha=0.5}$, and  ${e=g\in[0,0.5]}$. This shows the dependence on charges of the accelerating black holes.
}
\end{figure}

\begin{figure}[htp]
\begin{center}
\includegraphics[scale=0.53]{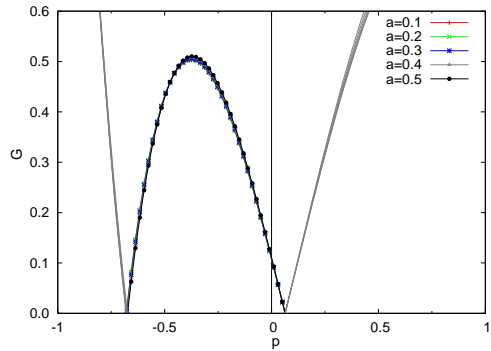}
\end{center}
\caption{\label{fig:graf21}%
The amplitude of gravitational radiation $|\mathcal{G}(p)|$ on a part of timelike $\mathcal{I}$ where ${\mathcal{Q}_{\mathcal{I}}<0}$, for ${\Lambda=-0.7}$, ${m=1}$, ${e=0.5=g}$, ${\alpha=0.5}$, and  ${a\in[0,0.5]}$. This shows that the dependence on rotation of the accelerating black holes is very weak.
}
\end{figure}

Again, generically the dependence on the acceleration $\alpha$ is  strong, the dependence on the charges $e$, $g$ is weaker (figure~\ref{fig:graf23}), and the dependence on the rotation $a$ is almost negligible (figure~\ref{fig:graf21}).


\subsection{Electromagnetic radiation}
\label{sec:amplitudeEMradiation}

\begin{figure}[htp]
\begin{center}
\includegraphics[scale=0.53]{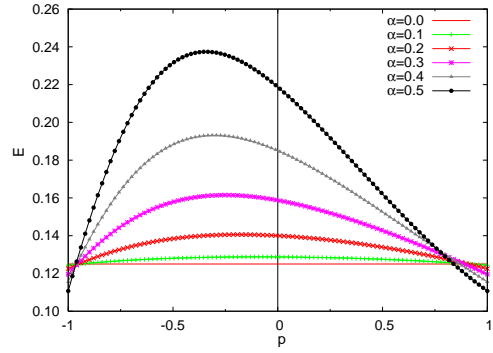}
\end{center}
\caption{\label{fig:grafEM1}%
The amplitude of electromagnetic radiation $\mathcal{E}(p)$ on a spacelike~$\mathcal{I}$, in which case ${\mathcal{Q}_{\mathcal{I}}<0}$ for any point $p$ of $\scri$. Here it is assumed that ${\Lambda=1}$, ${m=1}$, ${a=0.5}$, ${e=0.5=g}$, while the acceleration of the black holes is taken in the interval ${\alpha\in[0,0.5]}$. The horizontal line at ${\mathcal{E}=0.125}$ represents the uniform amplitude for non-accelerated Kerr--Newman--de-Sitter spacetime with ${\alpha=0}$.
}
\end{figure}

\begin{figure}[htp]
\begin{center}
\includegraphics[scale=0.53]{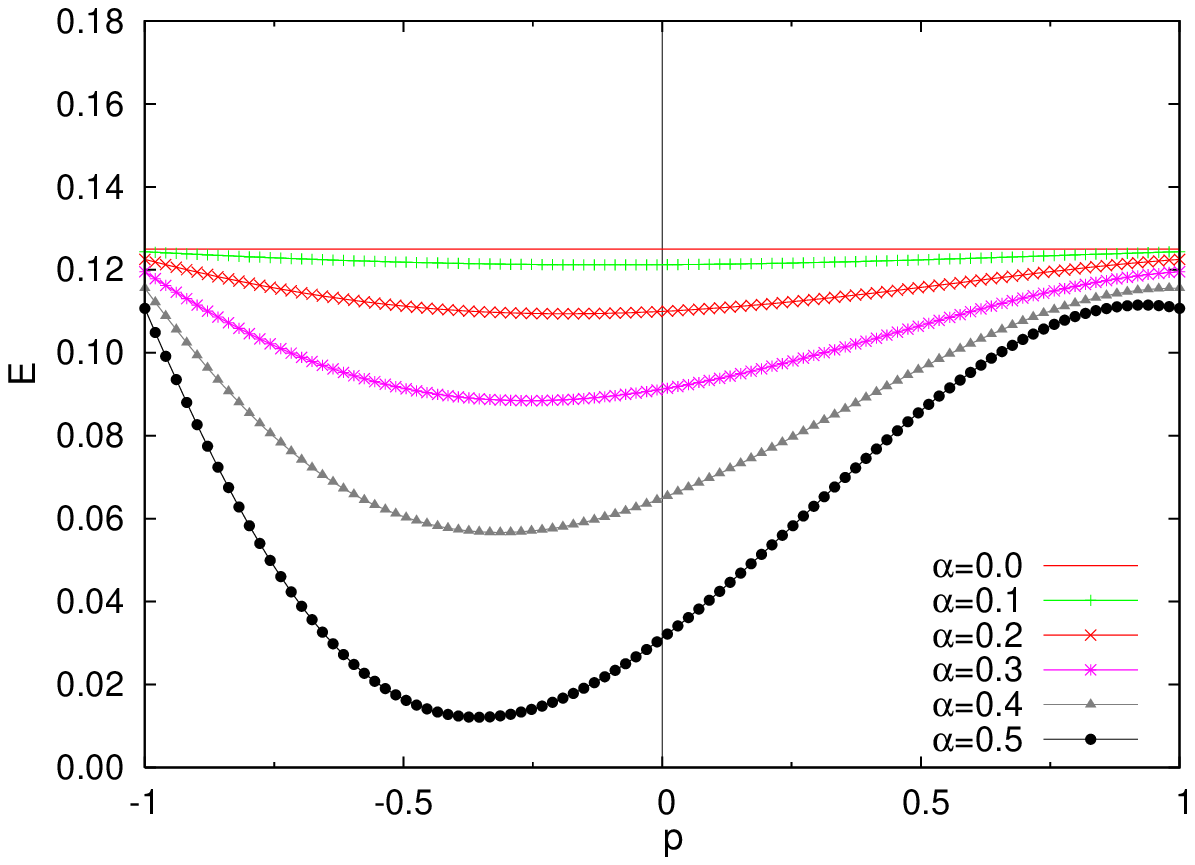}
\end{center}
\caption{\label{fig:grafEM6}%
The amplitude of electromagnetic radiation $\mathcal{E}(p)$ on a timelike $\mathcal{I}$ in the points where ${\mathcal{Q}_{\mathcal{I}}>0}$. Here ${\Lambda=-1}$, ${m=1}$, ${a=0.5}$, ${e=0.5=g}$, while the acceleration of the black holes is taken in the interval ${\alpha\in[0,0.5]}$.
}
\end{figure}

\begin{figure}[htp]
\begin{center}
\includegraphics[scale=0.53]{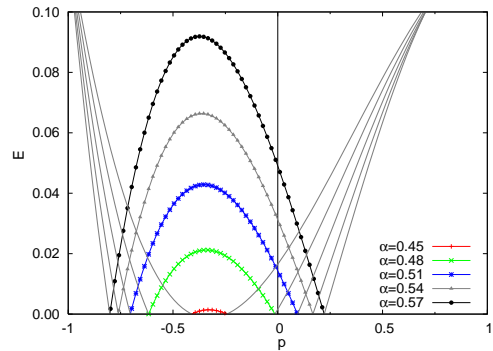}
\end{center}
\caption{\label{fig:grafEM11}%
The amplitude of electromagnetic radiation $\mathcal{E}(p)$ on a timelike $\mathcal{I}$ in the points where ${\mathcal{Q}_{\mathcal{I}}<0}$. Here ${\Lambda=-0.7}$, ${m=1}$, ${a=0.2}$, ${e=0.5=g}$, while the acceleration of the black holes is taken in the interval ${\alpha\in[0.45,0.57]}$.
}
\end{figure}

The amplitude of electromagnetic radiation ${{\mathcal E}(p)}$ is determined by expression (\ref{calEfinal}), in which the coordinate~$p$ labels all possible points on the (axially symmetric) conformal infinity~$\scri$. For a generic choice of specific parameters of accelerating and rotating black holes, this function is visualised as in figures~\ref{fig:grafEM1}--\ref{fig:grafEM11}. Figure~\ref{fig:grafEM1} applies to the case when ${\Lambda>0}$ while the remaining two figures apply when ${\Lambda<0}$. Specifically, figure~\ref{fig:grafEM6} describes the function ${{\mathcal E}(p)}$ in those regions of $\scri$ where ${\,\mathcal{Q}_{\scri} > 0}$, and figure~\ref{fig:grafEM11} in the points where ${\,\mathcal{Q}_{\scri} < 0}$. It can be seen that the behaviour of the electromagnetic amplitude ${{\mathcal E}(p)}$ is qualitatively similar to the behaviour of the gravitational amplitude ${{\mathcal G}(p)}$, cf. figures~\ref{fig:graf1}, \ref{fig:graf9} and \ref{fig:graf17}, respectively.


It should also be mentioned that the amplitude and directional structure of radiation, as described by expression (\ref{asyfieldEM1+final}), gives ($4\pi$-times) the magnitude of the Poynting vector in the parallelly transported interpretation frame. Moreover, it exactly reduces to results previously obtained for non-rotating black holes \cite{KrtousPodolsky:2003,PodolskyOrtaggioKrtous:2003}, and further for \emph{test} fields generated by uniformly accelerated charges in de~Sitter spacetime \cite{BicakKrtous:2002,BicakKrtous:2005} in the case when the mass of the sources becomes negligible.

\section{Conclusions}
In this contribution we analyzed in detail the asymptotic behaviour of gravitational and electromagnetic fields of black hole spacetimes which belong to the Pleba\'nski--Demia\'nski family of solutions. The amplitude and directional structure of radiation on conformal infinity~$\scri$ were evaluated in cases when the cosmological constant is non-vanishing, so that $\scri$ have either de Sitter-like or anti-de Sitter-like character. Although the accelerated systems studied do not seem to be astrophysically realistic (because their acceleration is caused by cosmic strings or struts attached to them along the axis of symmetry), the results obtained elucidate some principal geometric and physical properties of this class of exact radiative spacetimes.

In particular, we found explicit relations between the parameters that characterize the sources, namely their mass, electric and magnetic charges, NUT parameter, rotational parameter, and acceleration, and the asymptotic radiative properties. For example, it was demonstrated that radiation, described by specific components of the Weyl and Maxwell tensors in the parallelly transported interpretation frame near conformal infinity, is not distributed uniformly on $\scri$ when the black holes accelerate. Also, both the gravitational amplitude ${{\mathcal G}(p)}$ and the electromagnetic amplitude ${{\mathcal E}(p)}$ of radiation on~$\scri$ depend strongly on the acceleration of the black holes, weakly on their charges, and very weakly on their rotation. Since previously only non-rotating black holes (described by the C-metric) were studied, this is a new and rather surprising result. 

As already mentioned above, there are fundamental differences concerning the asymptotic structure of radiation in spacetimes with and without a cosmological constant. Since the geometric nature of the corresponding conformal infinities is different it is not straightforward to perform the limit ${\Lambda\to 0}$. Indeed, the amplitudes ${{\mathcal G}(p)}$ and  ${{\mathcal E}(p)}$, given by (\ref{calGfinal}) and (\ref{calEfinal}), diverge in this limit. Nevertheless, in the case ${\Lambda=0}$ it is possible to evaluate \emph{directly} the dominant (radiative) components of the fields in an appropriate interpretation frame. It follows from (\ref{nice}) that ${\mathcal{Q}<0}$ near the conformal infinity (which now has a null character), so that the region must be non-stationary. Using then the reference frame
${{\bf t}_{\rm o}={\bf t}_{\rm s}, {\bf q}_{\rm o}={\bf r}_{\rm s}, {\bf r}_{\rm o}={\bf q}_{\rm s}, {\bf s}_{\rm o}={\bf s}_{\rm s}}$, where ${{\bf t}_{\rm s}, {\bf q}_{\rm s}, {\bf r}_{\rm s}, {\bf s}_{\rm s}}$ are given by (\ref{sp+}), it turns out that the gravitational and electromagnetic fields behave asymptotically as 
${|{\Psi }_4^{\rm i}|  \approx 3 |{\Psi }_{2\ast}^{\rm s}|\,|\eta|^{-1}}$ and
${|{\Phi }_2^{\rm i}|  \approx \sqrt{2}\, |{\Phi }_{1\ast}^{\rm s}|\,|\eta|^{-1}}$, where the coefficients are given by (\ref{weyl1alterspec}) and (\ref{elmag1alterspec}), i.e., 
${|{\Psi }_{2\ast}^{\rm s}|=\sqrt{\mathcal{D}(p)}\,(1+\alpha^2\omega^2p^4)^{-\frac{3}{2}}}$ and ${|{\Phi }_{1\ast}^{\rm s}|=\frac{1}{2}\sqrt{e^2+g^2}\,(1+\alpha^2\omega^2p^4)^{-1}}$, with ${\mathcal{D}(p)}$ defined by (\ref{Dobecne}). Compared to (\ref{asyfield1+final}) and (\ref{asyfieldEM1+final}), there is obviously no directional structure when ${\Lambda=0}$ (the fields are the same for all null geodesics approaching a given point at null conformal infinity). Also, with the above normalisation, the dependence on $p$ (which labels all possible points at $\scri$) is simpler and the amplitudes of gravitational and electromagnetic radiation are finite.

\section*{Acknowledgments}
This work was supported by the grant GA\v{C}R~202/08/0187 and by the Czech Ministry of Education under the projects MSM0021610860 and LC06014.

\end{document}